# Optical Spin Sensing and Metamagnetic Phase Control in the 2D Van der Waals Magnet Yb$^{3+}$-Doped CrPS$_4$

Jacob T. Baillie,[1] Kimo Pressler,[1] Nick J. Adams,[1] Faris Horani,[1] Thom J. Snoeren,[1] Rémi Beaulac,[2] and Daniel R. Gamelin[1,*]

[1]*Department of Chemistry, University of Washington, Seattle, WA 98195*
[2]*Department of Chemistry, Swarthmore College, Swarthmore, PA 19081*

*Correspondence to: gamelin@chem.washington.edu

**Abstract:** The emergence of two-dimensional magnets within the van der Waals toolkit has introduced unprecedented opportunities to develop ultrathin spintronic technologies. Strong coupling between spin and optical properties in such materials can further enable novel spin-photonic capabilities of both fundamental and technological interest. Here, we investigate the optical and spin properties of the air-stable, layered A-type antiferromagnet chromium thiophosphate (CrPS$_4$) when doped with Yb$^{3+}$. We show that the collective spin properties of CrPS$_4$ are encoded in the sharp *f-f* luminescence of isolated Yb$^{3+}$ dopants *via* strong magnetic superexchange coupling between the two, and that spontaneous magnetic ordering in CrPS$_4$ induces large exchange splittings in the narrow Yb$^{3+}$ *f-f* photoluminescence features below $T_N$. Spin reorientation in CrPS$_4$ *via* a "spin-flop" metamagnetic transition modulates the Yb$^{3+}$ *f-f* luminescence energies and exchange splittings. This pronounced link between spin and optical properties enables the demonstration of optically driven spin-flop transitions in CrPS$_4$.

**Main Text**

The layered van der Waals (vdW) magnet, CrPS$_4$, is an emerging material showing exotic magneto-electronic properties including voltage-modulated electron tunneling,[1-3] spin-polarized transport,[4,5] and magnetoconductance.[5,6] Its in-plane anisotropy allows anisotropic tunneling magnetoresistance,[7-9] anisotropic magnon transport[10] and magnon chirality switching,[11] anisotropic linear- and nonlinear optical responses,[12] and it shows saturable absorption.[13] CrPS$_4$ also exhibits rich, structured photoluminescence (PL),[14-17] but unlike other 2D magnetic materials this PL shows almost no response to applied magnetic fields.[14] Recently, magnetic circularly polarized luminescence (MCPL) was shown to track the out-of-plane magnetization of CrPS$_4$,[18] but the polarization response is weak.

The CrPS$_4$ lattice consists of highly anisotropic 2D layers involving parallel 1D chains of edge-sharing [CrS$_6$]$^{3-}$ octahedra separated by S-P-S bridges, alternating between two crystallographically distinct Cr$^{3+}$ sites.[19] No other MPS$_4$ or MPS$_3$ compound shares this structure.[20,21] Magnetic susceptibility[22,23] and neutron diffraction[24] have established that each monolayer orders ferromagnetically with its Cr$^{3+}$ moments nearly perpendicular to the 2D plane, and neighboring monolayers couple antiferromagnetically below a Néel temperature ($T_N$) of ~36 K, making bulk CrPS$_4$ an A-type antiferromagnet. An abrupt metamagnetic spin-flop transition (SFT) is observed with ***B*** || *c* (out-of-plane) of ~1 T at 5 K, reorienting all spins into the 2D plane while retaining antiferromagnetic interlayer coupling. Higher magnetic fields progressively cant the Cr$^{3+}$ spins back out of plane until full ferromagnetic (FM) alignment is reached at ~8 T. Exfoliation down to mono- or few-layer structures reveals layer-dependent magnetism, with even- and odd-layer structures alternating between zero and non-zero magnetic moments, respectively.[25]

Lanthanides doped into crystals have attracted broad attention as phosphors, laser gain media, and more recently, as spin-bearing optical impurities for potential quantum information applications. Screening of their spectroscopically active 4*f* valence electrons imparts trivalent lanthanides with narrow optical linewidths, high PL quantum yields, and long spin- and optical-coherence times,[26,27] enabling optical quantum memory,[28-30] microwave-to-optical quantum transduction,[31] and coherent manipulation and optical readout of individual spins.[32] In many instances, magnetic fields are applied to split spin degeneracies for optical spin initialization. Whereas CrPS$_4$ doping remains wholly unexplored, lanthanide (Yb$^{3+}$) doping of other 2D magnets (CrI$_3$, CrBr$_3$) was recently established as a route to novel spin-photonic properties,[33,34] including spontaneous splittings of Yb$^{3+}$ Kramers spin degeneracies *via* superexchange coupling with the surrounding lattice. This work prompts the investigation of 2D magnets showing exotic metamagnetic properties, such as CrPS$_4$.

Here we demonstrate that doping with even small amounts (<1% *vs* Cr$^{3+}$) of Yb$^{3+}$ quenches the native CrPS$_4$ PL and generates strong sensitized Yb$^{3+}$ *f-f* PL. Low-temperature PL measurements at zero magnetic field show complex fine structure attributable to Yb$^{3+}$ substituting at the two Cr$^{3+}$ lattice sites, each showing well-resolved spin splittings due to superexchange coupling with surrounding Cr$^{3+}$ ions. These Yb$^{3+}$ dopants show an extraordinarily strong optical response to the CrPS$_4$ spin-flop transition, with large PL energy shifts and effective *g* values reaching ~170. Exploiting this strong coupling between optical and magnetic properties, we demonstrate the use of $Yb^{3+}$ PL to detect and quantify photothermally driven $CrPS_4$ spin-flop transitions. These results expand upon the spin-photonic functionality known for lanthanides and the optical functionalities known for 2D magnets, with interesting potential ramifications for photo-spintronics and quantum information sciences.

**Results and Analysis**

$CrPS_4$ and $Yb^{3+}$-doped $CrPS_4$ ($Yb^{3+}$:$CrPS_4$) crystals were grown by chemical vapor transport, and $Yb^{3+}$ concentrations were measured by inductively coupled plasma mass spectrometry (ICP-MS, see Methods). Figure 1a shows X-ray diffraction (XRD) data collected on a single-crystal flake of 0.2% $Yb^{3+}$:$CrPS_4$ using a powder diffractometer, alongside a reference powder pattern for undoped $CrPS_4$. The XRD peak positions of the doped sample align well with those of the reference. The (00$l$) peaks are relatively more intense than in the reference pattern, reflecting orientation of the flake with its $c$-axis normal to the substrate. Figure 1b shows the Raman spectrum collected from a 0.2% $Yb^{3+}$:$CrPS_4$ flake, with red and blue labels indicating literature peak assignments.[35,36] The inset shows a photograph of a representative $Yb^{3+}$:$CrPS_4$ flake. At this low doping level, the XRD, Raman, and crystal morphologies of $CrPS_4$ and $Yb^{3+}$:$CrPS_4$ are essentially indistinguishable.

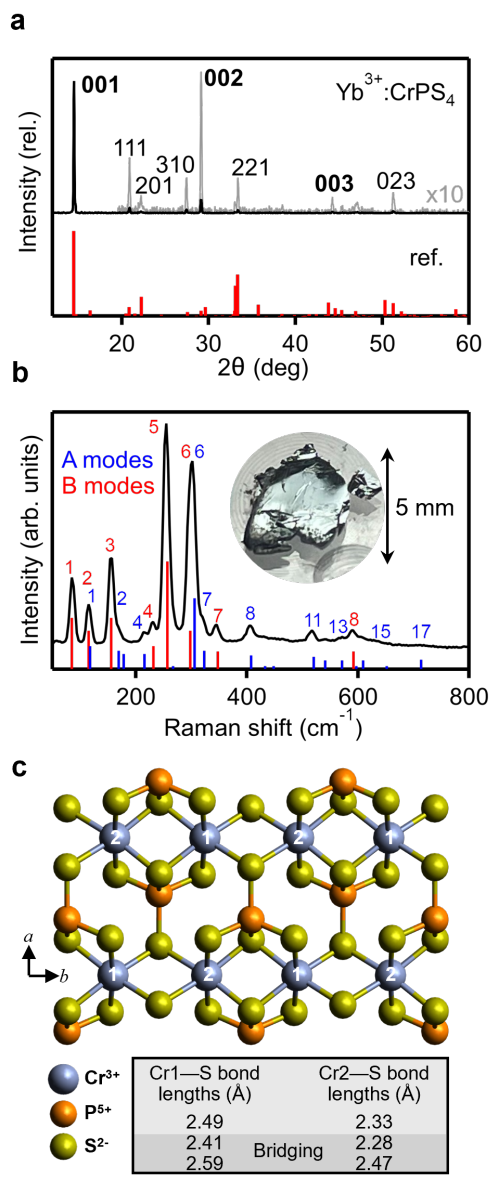

**Figure 1.** **(a)** XRD data collected on a 0.2% $Yb^{3+}$:$CrPS_4$ single crystal using a powder diffractometer. A reference powder diffraction pattern for $CrPS_4$ is shown for comparison (red, ICSD Coll. Code 937). The experimental (00*l*) peaks are disproportionately strong, consistent with orientation of the 2D planes parallel to the substrate. **(b)** Raman spectrum collected from a 0.2% $Yb^{3+}$:$CrPS_4$ single crystal using unpolarized 532 nm excitation. Peaks are assigned based on refs. 35,36. Intensities of the reference Raman lines are scaled arbitrarily to guide the eye. The inset shows a photograph of a representative $Yb^{3+}$:$CrPS_4$ flake. **(c)** The 4 K structure of $CrPS_4$, showing alternation between two pseudooctahedral $Cr^{3+}$ sites (Cr1 and Cr2) with different Cr-S bond lengths (from ref. 24).

The low-temperature PL spectrum of undoped CrPS$_4$ has been described in depth previously.[14-16,37] This spectrum shows multiple evenly spaced peaks between 10500 and 11000 cm$^{-1}$ that have been attributed to vibronic coupling, as well as additional rich fine structure at the lowest temperatures.[14,15] This spectrum also typically shows a broad trap feature near 9000 cm$^{-1}$, although as emphasized previously,[37] many reports do not adequately account for this low-energy emission.

Figure 2a compares the PL spectrum of a 0.2% Yb$^{3+}$:CrPS$_4$ crystal measured at 36 K with that of an undoped CrPS$_4$ crystal measured under the same conditions. This temperature was chosen for simplicity because the additional CrPS$_4$ PL fine structure observed at lower temperatures is unrelated. In the PL spectrum of 0.2% Yb$^{3+}$:CrPS$_4$, the signature features of undoped CrPS$_4$ are replaced by a series of narrow peaks between 9000 and 10100 cm$^{-1}$ that are characteristic of Yb$^{3+}$ *f-f* luminescence. Replotting the high-energy part of the Yb$^{3+}$:CrPS$_4$ PL spectrum with intensities magnified 2000x reveals that the native PL of CrPS$_4$ is still present, just very weak. Integration shows that >99.9% of the PL intensity now comes from Yb$^{3+}$-centered transitions. With only one Yb$^{3+}$ impurity per ~500 Cr$^{3+}$ ions at 0.2% doping, this result indicates extensive exciton diffusion within CrPS$_4$.

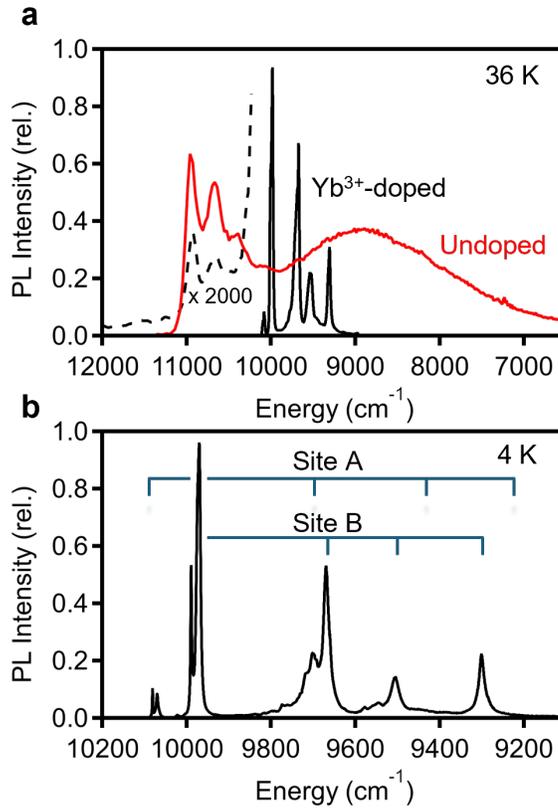

**Figure 2. (a)** 36 K PL spectrum of CrPS$_4$ (red) and 0.2% Yb$^{3+}$:CrPS$_4$ (black) single crystals. The dashed curve plots the same PL spectrum of 0.2% Yb$^{3+}$:CrPS$_4$ scaled

up 2000x. **(b)** 4 K PL spectrum of the $Yb^{3+}$ $^2F_{5/2} \rightarrow {}^2F_{7/2}$ PL from the 0.2% $Yb^{3+}$:$CrPS_4$ crystal plotted on an expanded scale. The brackets denote the primary features identified for two distinct $Yb^{3+}$ doping sites, A and B, using site-selective photoexcitation. All 0.2% $Yb^{3+}$:$CrPS_4$ PL data were measured using unpolarized CW 632.8 nm photoexcitation at 5 mW/cm$^2$. The $CrPS_4$ spectrum was measured using unpolarized pulsed 470 nm photoexcitation (6.3 nJ per 3 ns pulse, 40 MHz, average power density of 2.1 W/cm$^2$, $\vec{E} \perp c$).

The $^2F$ free-ion term of $Yb^{3+}$ ($4f^{13}$ configuration) is split into $^2F_{5/2}$ and $^2F_{7/2}$ terms by strong spin-orbit coupling (~10000 cm$^{-1}$), and each of these is split by the low-symmetry crystal field in $CrPS_4$ (~500 cm$^{-1}$) to yield a Kramers-doublet ground state, denoted $^2F_{7/2}(0)$, and a series of Kramers doublet excited states. Figure 2b plots the 4 K PL spectrum of the $Yb^{3+}$-doped sample from Figure 2a on an expanded energy scale. This spectrum is complex, showing more peaks than anticipated for a distorted octahedral $Yb^{3+}$ ion. Using site-selective laser excitation of $Yb^{3+}$, two species could be identified in the PL (A and B, Figures 2b, S1-S2). These species are assigned to substitutional $Yb^{3+}$ in the two distinct $Cr^{3+}$ lattice sites of $CrPS_4$. Site B contributes most of the PL intensity, with Site A PL ~10x weaker (Figure 2b), suggesting preferential B-site doping. $Yb^{3+}$ is much larger than $Cr^{3+}$; the smaller crystal-field splitting for $Yb^{3+}$ in Site B suggests association of this site with the less-constrained Cr1 lattice site illustrated in Figure 1c. These structural differences may also contribute to the shorter 4 K PL decay time of $Yb^{3+}$ in Site A (51 μs) compared to Site B (82 μs) (Figure S1c). Even accounting for these two sites, the spectrum shows more peaks than expected. In particular, *two* peaks are observed at the highest-energy electronic origins of both A- and B-site $Yb^{3+}$ PL (centered at 10075 and 9980 cm$^{-1}$ at 4 K, respectively), suggesting that their ground-state Kramers degeneracies are lifted by coupling with the surrounding lattice. Unlike in most materials, $Yb^{3+}$ dopants in $CrPS_4$ thus do not possess time-reversal symmetry at 4 K, despite the bulk antiferromagnetism.

To illustrate, Figure 3 summarizes variable-temperature PL data collected for the 0.2% $Yb^{3+}$:$CrPS_4$ crystal from Figure 2, focusing on the first electronic origins of each $Yb^{3+}$ site. Figure 3a shows PL spectra for $Yb^{3+}$ in Site A collected from 4 to 75 K and plotted spectrally and as a heat map (see Figure S3 for complete data set). At 4 K, two peaks denoted $α$ and $β$ are observed. With increasing temperature, both peaks broaden and $β$ shifts to higher energy. Concurrently, a pair of hot bands (peaks $α^H$ and $β^H$) emerges at higher energy, gaining intensity and shifting to lower energy with increasing temperature. Peaks $α$ and $β$ are assigned as PL transitions to the two spin components of an exchange-split ground $^2F_{7/2}(0)$ Kramers doublet (denoted $0^+$ and $0^-$), as summarized in the inset to Figure 3a. Likewise, peaks $\alpha^H$ and $\beta^H$ are assigned to PL originating from the thermally populated upper level of the exchange-split $^2F_{5/2}(0')$ excited-state doublet (denoted $0'^+$). The temperature-dependence of these PL intensities (Figure S4) supports these assignments. Figure 3 shows that the splitting energies between $0^+$ and $0^-$ (denoted $\Delta E$) and between $0'^+$ and $0'^-$ (denoted $\Delta E'$) decrease as the temperature increases from 4 K to the CrPS$_4$ Néel temperature, where they abruptly stop being temperature dependent. This result demonstrates a strong sensitivity of the Yb$^{3+}$ PL to the spin correlations that emerge spontaneously below $T_N$ in CrPS$_4$. Figure 3b plots energy differences between select peaks in Figure 3a. Per the peak assignments, $E(\alpha) - E(\beta)$ and $E(\alpha^H) - E(\beta^H)$ both equal $\Delta E$, whereas $E(\alpha^H) - E(\alpha)$ and $E(\beta^H) - E(\beta)$ both equal $\Delta E'$. Because no external magnetic field is used, these splittings are solely due to magnetic exchange coupling, such that $\Delta E = \Delta E_{exch}$ and $\Delta E' = \Delta E'_{exch}$. Figure 3b plots the temperature dependence of $\Delta E_{exch}$ and $\Delta E'_{exch}$. $\Delta E_{exch} = 11$ cm$^{-1}$ at 4 K and decreases with increasing temperature until $T_N$ before becoming independent of temperature at a value of 7.5 cm$^{-1}$. Similarly, $\Delta E'_{exch} = 27$ cm$^{-1}$ at 4 K and decreases with temperature until plateauing at 19 cm$^{-1}$ above $T_N$. As also seen in Yb$^{3+}$:CrBr$_3$,[34] $\Delta E_{exch}$ remains non-zero even in the paramagnetic regime because the dominant Yb$^{3+}$-Cr$^{3+}$ exchange interactions are short-range and even random spin disorder leads to net Cr$^{3+}$ spin around Yb$^{3+}$ dopants. Above ~75 K, the Yb$^{3+}$ PL intensity and decay time both drop, indicating thermally activated nonradiative losses (Figure S5).

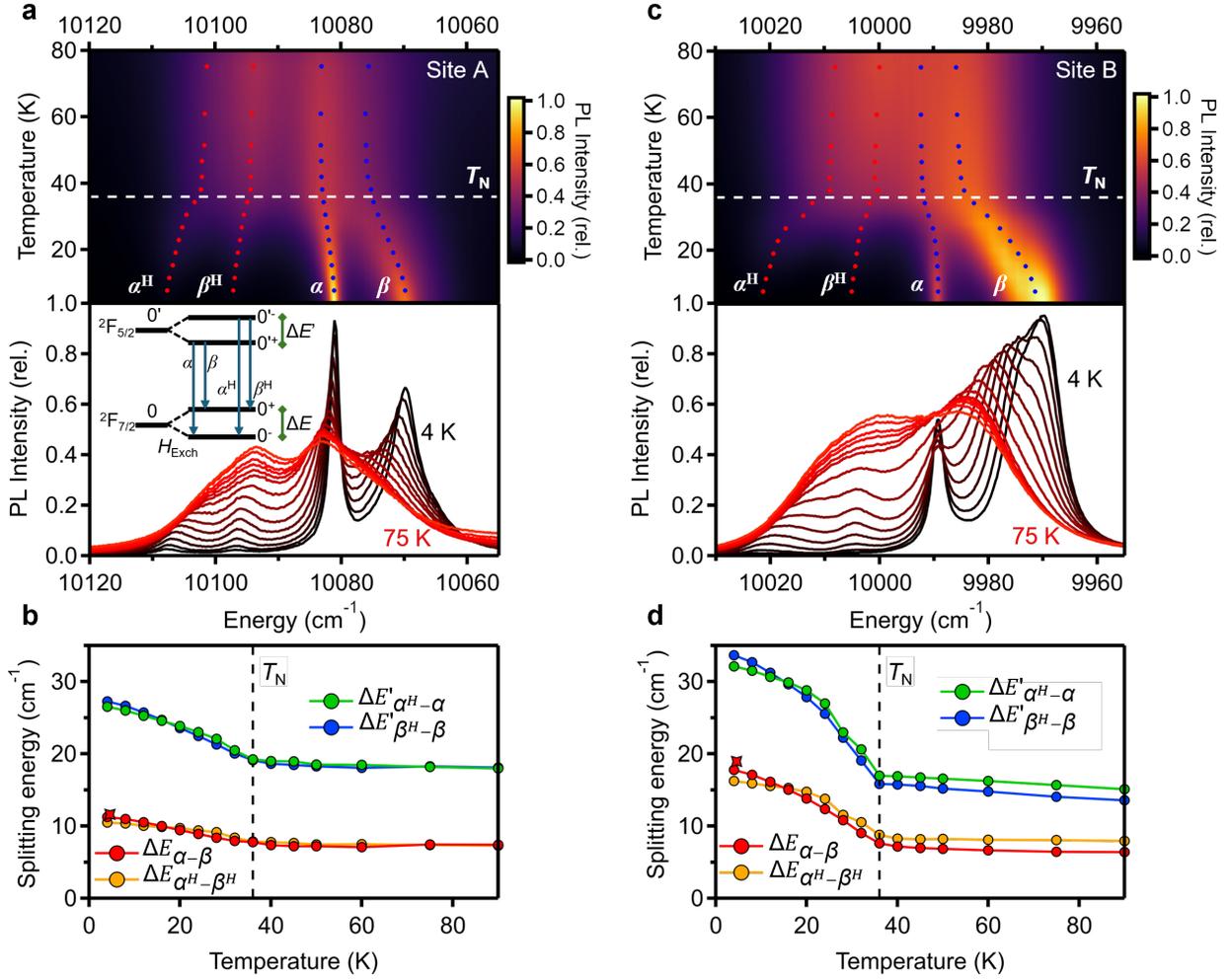

**Figure 3. (a)** Top: False-color plot of the PL intensity from 4 to 75 K for the 0' → 0 PL peaks of $Yb^{3+}$ in Site A of a 0.2% $Yb^{3+}$:$CrPS_4$ single crystal. The overlayed data points mark the energies of $\alpha^H$ and $\beta^H$ (red) and $\alpha$ and $\beta$ (blue) peaks, determined by deconvolution (Figure S6). The dashed white line indicates $T_N$ = 36 K for $CrPS_4$. Bottom: The PL spectra used to make the false-color plot above. The inset illustrates the specific PL transitions observed between exchange-split $Yb^{3+}$ states at the 0' → 0 electronic origin. **(b)** Temperature dependence of $\Delta E_{exch}$ and $\Delta E'_{exch}$ for $Yb^{3+}$ at Site A, determined from the data in panel (a) as described in the main text. **(c)** Same as panel (a) but for $Yb^{3+}$ in Site B. **(d)** Same as panel (b) but for Site B. The red x symbols at 4.6 K in panels (b) and (d) show data points collected in a helium immersion cryostat (Figures 4, S8), which agree well with the data collected in a closed-cycle cryostat and confirm the accuracy of the sample temperature readings in the latter. All PL data were measured using unpolarized CW 632.8 nm photoexcitation at 5 mW/cm² with $\vec{E} \perp c$.

Figures 3c,d show parallel PL data for the first electronic origin of $Yb^{3+}$ in Site B. The same trends in peak energies, hot-band intensities, and peak broadening are all apparent. Like in Site A, $\Delta E_{exch}$ and $\Delta E'_{exch}$ also decrease from their maximum values (17 and 33 cm$^{-1}$, respectively) at 4 K until $T_N$ and then become temperature-independent (at $\Delta E_{exch}$ = 7.5 and 16 cm$^{-1}$, respectively). Site B shows larger exchange splittings than Site A despite its smaller $^2F_{7/2}$ crystal-field splitting, and its discontinuity in $\Delta E^{(\prime)}_{exch}$ at $T_N$ is more abrupt. Because of the similarity between Site A and B data, the rest of this manuscript focuses on characterization of the majority Site B, with additional Site A data provided as Supplementary Material. Overall, the data in Figure 3 demonstrate that the PL of the $Yb^{3+}$ impurities is strongly correlated with the spin properties of the $CrPS_4$ lattice, indicating both optical and magnetic integration of this spin-bearing point defect into the material.

To probe this magnetic integration in more detail, PL measurements were performed under applied magnetic fields ($\bm{B}\|c$). Figure 4 summarizes data for the $\alpha$ and $\beta$ peaks of Site B in 0.02% $Yb^{3+}$:$CrPS_4$ (see Figure S7 for the full data set, Figure S8 for Site A data). Figure 4a plots 4.6 K PL spectra collected at select field strengths between 0 and 6 T. Very little changes at low field until around 0.93 T, where the $\alpha$ and $\beta$ peaks broaden and shift to higher energy. By 1.05 T, both peaks have shifted by about +10 cm$^{-1}$ and structure appears in the $\alpha$ peak (Figure S9, S10). At 2 T, the peak broadening has mostly subsided again. The heat map in Figure 4b highlights the abruptness of these spectral changes at low magnetic fields. This unusual field dependence is attributable to the metamagnetic spin-flop transition (SFT) of $CrPS_4$, in which $Cr^{3+}$ spin vectors concertedly rotate from out-of-plane to in-plane orientations while retaining the interlayer antiferromagnetic coupling (A-AFM → cAFM). This spin reorientation affects the $Yb^{3+}$ PL energies because of the strong prolate anisotropy of $Yb^{3+}$. The SFT field ($B_{SFT}$) and its abruptness observed optically are both consistent with magnetic susceptibility results.[22,38] The peak broadening and resolved structure at intermediate fields reveal the presence of magnetic domains with slightly different $B_{SFT}$ values. From 2 to 6 T, $\alpha$ and $\beta$ gradually shift to lower energy and sharpen further as $Cr^{3+}$ spins cant out of plane, and by 6 T the spectrum again bears similarity to the 0 T spectrum. These results highlight the high sensitivity of $Yb^{3+}$ PL as an optical probe of $CrPS_4$ magnetism.

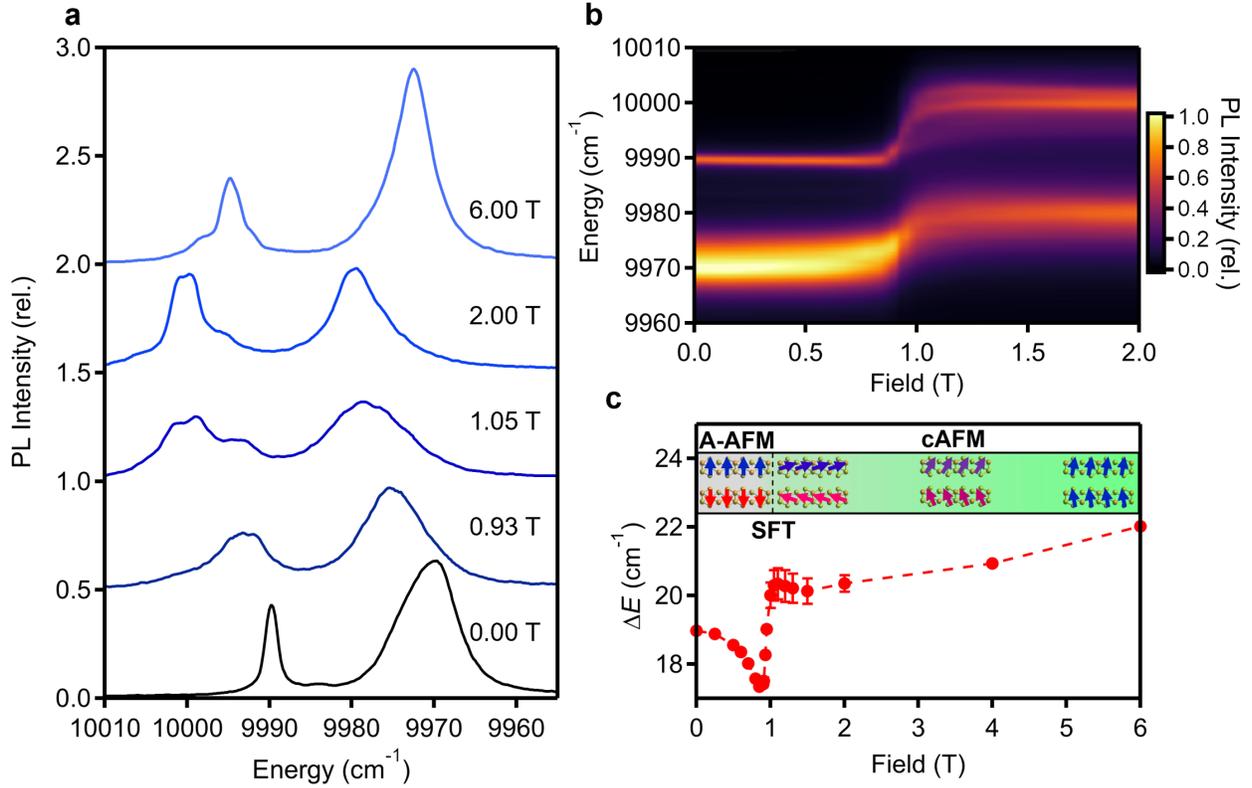

**Figure 4.** (a) 4.6 K variable-field PL spectra in the region of the α and β peaks for $Yb^{3+}$ dopants at Site B of an oriented 0.02% $Yb^{3+}$:$CrPS_4$ single crystal. (b) False-color plot of the PL intensities *vs* magnetic field strength in the region of the α and β peaks for Site B. (c) Plot of $\Delta E$ *vs* magnetic field strength for Site B. Determination of the peak energies is described in Figures S9 and S10. The inset illustrates the $Cr^{3+}$ spin order *vs* magnetic field strength. For all panels, data were measured at 4.6 K using unpolarized CW 632.8 nm photoexcitation at 5 mW/cm² with $\vec{E} \perp c$ and $\boldsymbol{B} \parallel c$.

Applying a magnetic field introduces an additional Zeeman term to each doublet's energy splitting, such that $\Delta E = \Delta E_{exch} + \Delta E_Z$ (and likewise for $\Delta E'$). Figure 4c plots $\Delta E$ *vs* $\boldsymbol{B}$ for the ground state of $Yb^{3+}$ in Site B, measured at 4.6 K. These data show a dramatic derivative-shaped response at $B_{SFT}$. The initial dip in $\Delta E$ comes from the β peak shifting at slightly lower $\boldsymbol{B}$ than the α peak. $\Delta E$ then jumps abruptly by ~+3 cm⁻¹ at $B_{SFT}$. Above 1 T, $\Delta E$ changes very slowly again up to 6 T. $\partial(\Delta E)/\partial\boldsymbol{B}$ maximizes at +37 cm⁻¹/T at $B_{SFT}$, corresponding to an effective g value of $g_{eff}$ = +79 and indicating a very high sensitivity to the external magnetic field.

Although the excited-state splitting is not measurable by PL at 4.6 K, the $α^H$ and $β^H$ hot bands observed at 16 K allow analysis of both $\Delta E'$ and $\Delta E$ at this temperature (Figure S11). In these data, $\Delta E'$ jumps by ca. -21 cm⁻¹ at $B_{SFT}$, compared to $\Delta E \approx +2.5$ cm⁻¹ for the ground state at the same temperature, with $(\partial(\Delta E')/\partial B)_{max}$ = -80 cm$^{-1}$/T at $B_{SFT}$ ($g_{eff}$ = -172) compared to $(\partial(\Delta E)/\partial B)_{max}$ = +17 cm$^{-1}$/T ($g_{eff}$ = +40) for the ground state. The luminescent excited state is thus ca. 5x more sensitive to the SFT than the ground state. The sign change of the exchange splitting reflects Yb$^{3+}$ having oppositely signed spin components along the total angular momentum directions of its $^2F_{7/2}(0)$ and $^2F_{5/2}(0')$ states.[34,39] A comparison of energy levels and splittings at 4.6 and 16 K is compiled in Figure S13.

These results represent a significant improvement in sensitivity over prior optical probes of CrPS$_4$ magnetism. CrPS$_4$ PL energies appear insensitive to the SFT,[14] but PL polarizations allow detection of this phase transition:[18] the polarization ratio $\rho = (\sigma^- - \sigma^+)/(\sigma^- + \sigma^+)$, where $\sigma^-$ and $\sigma^+$ are the intensities of left and right circularly polarized PL, respectively, increases from 0 to ~2% crossing $B_{SFT}$. Variable-temperature Raman data at 0 T may also show a small inflection at $T_N$ but 5 K data show no discernible inflection at $B_{SFT}$.[18] In contrast, the $\alpha$ and $\beta$ PL peaks of Yb$^{3+}$ dopants in CrPS$_4$ shift in energy by over one linewidth crossing $B_{SFT}$, providing extremely high sensitivity. To illustrate, Figure 5a plots the PL energy of the combined $\alpha$ and $\beta$ center of mass *vs* $B$ relative to its zero-field value. This energy shift and the value of $B_{SFT}$ both decrease as temperature increases from 4 to 36 K, above which no SFT is detected. Figure 5b summarizes these changes in a magnetic phase diagram, showing a temperature-dependent phase boundary that disappears above $T_N \approx 36$ K. This optical result agrees well with analogous phase diagrams obtained from magnetic susceptibility measurements.[22,23]

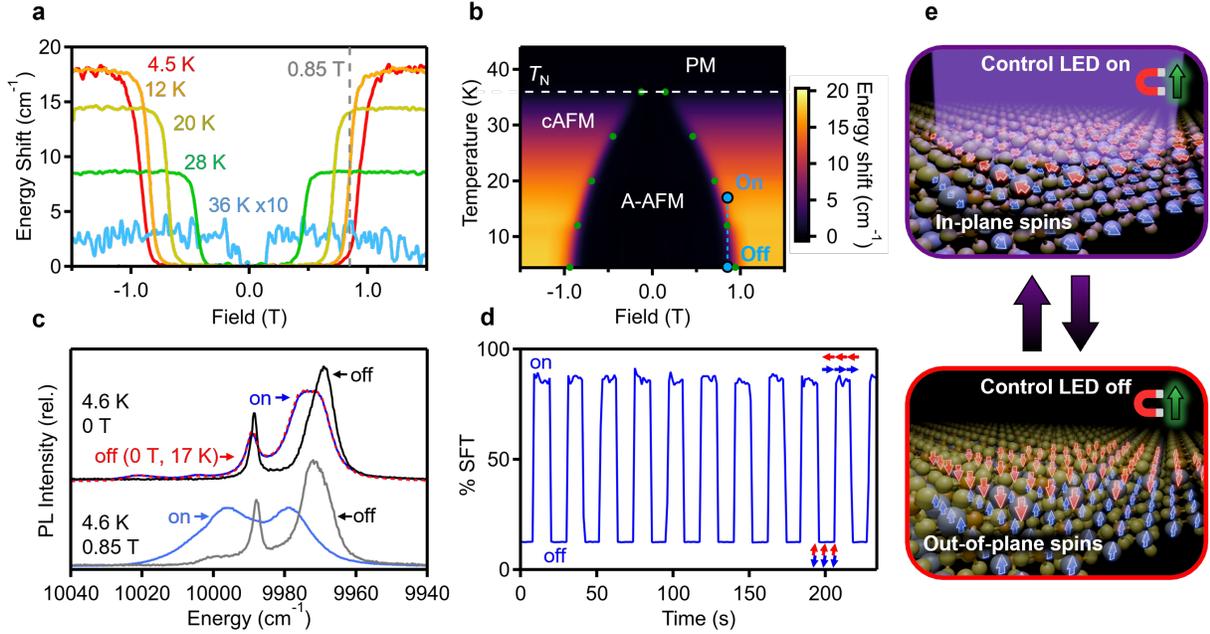

**Figure 5. (a)** Magnetic-field dependence of Site B $Yb^{3+}$ $0' \to 0$ PL energies plotted relative to their zero-field energies, measured at various temperatures. The energies shown here represent the centers of gravity for the combined $\alpha$ and $\beta$ peaks (9940 – 10040 cm$^{-1}$). Data were collected using 5 mW/cm$^2$ of unpolarized CW 632.8 nm photoexcitation with and without co-excitation from an unpolarized 405 nm, 140 W/cm$^2$ "control" LED. **(b)** False-color plot of the data in panel (a). The green data points indicate $B_{SFT}$ values at each experimental temperature. Interpolation to create the heatmap was performed using the trendlines shown in Figure S14. The horizontal dashed white line denotes $T_N$ of CrPS$_4$. **(c)** PL spectra with and without control LED co-excitation. Black: 0 T, 4.6 K, no control LED. Dark blue: 0 T, 4.6 K, with control LED. Dashed red: 0 T, 17 K, no control LED. Gray: 0.85 T, 4.6 K, no control LED. Light blue: 0.85 T, 4.6 K, with control LED. **(d)** Reversible optically driven SFT obtained by modulation of the control LED in 10 s on and off intervals, measured probing $Yb^{3+}$ PL with the cryostat at 4.6 K. % SFT is defined as the fractional transition from low- to high-field limits of the SFT transition, see Figure S16a. **(e)** Schematic summary of the results from panel (d), showing optical spin-reorientation in CrPS$_4$ by the control LED. All panels are for an oriented 0.02% $Yb^{3+}$:CrPS$_4$ single crystal (same as Figure 4) with $\vec{E} \perp c$ and $\boldsymbol{B} \parallel c$.

We now use this sensitivity to demonstrate *control* of CrPS$_4$ magnetism using light. Previously, picosecond photothermal demagnetization has been achieved across the series of analogous MPS$_3$ antiferromagnets (M = $Co^{2+}$, $Mn^{2+}$, $Fe^{2+}$, $Ni^{2+}$),[40-43] demonstrating effective phonon-magnon coupling in such compounds. Photoexcitation above bandgap causes the phonon temperature to increase relative to that of the magnons because of phonon-assisted absorption and nonradiative relaxation. Phonon-magnon coupling then raises the spin temperature and destroys the $MPS_3$ magnetic ordering. Such optically driven demagnetization is attractive for both fundamental studies of phonon-magnon interactions and for extremely fast manipulation of correlated spins in layered quantum materials. We therefore sought to demonstrate an analogous optically driven *metamagnetic* phase transition in $CrPS_4$, probed using the high sensitivity of the $Yb^{3+}$ PL.

Figure 5c plots $Yb^{3+}$:$CrPS_4$ PL spectra obtained with and without an additional "control" CW light-emitting diode (LED) exciting the sample at 405 nm. At 4.6 K and zero magnetic field, introducing the control LED simply broadens the spectrum without shifting the first peak's energy, consistent with lattice heating. The internal temperature with the control LED on is determined to be ~17 K by comparison with variable-temperature PL spectra collected without the control LED (Figures 5c, S15). Introducing a 0.85 T magnetic field yields markedly different results: the control LED now shifts both peaks to substantially higher energies, a signature of the SFT (Figure 4). Analysis of this shift (Figure S16a) indicates the control LED causes a jump from 12.5 to 87% of the full field-induced SFT, and analyses of $\Delta E$, $E(\alpha)$, and $E(\beta)$ yield similar values (Figure S16). This result is quantitatively consistent with the change in magnetization between 5 and 17 K measured at 0.85 T.[22,23] Neutron diffraction[22] shows out-of-plane spins at 0 T and 1.7 K oriented $\theta$ = 9.6° off of the crystallographic *c*-axis, and magnetic susceptibility data[22,23] suggest $\theta \approx 85°$ at 0.85 T and 17 K. The results in Figure 5c thus show nearly complete out-of-plane to in-plane spin reorientation driven by the control LED. Figure 5d further demonstrates that square-wave modulation of the control LED reversibly switches the $CrPS_4$ spin ordering between A-AFM and cAFM phases, as summarized schematically in Figure 5e and highlighted by the blue circles at 0.85 T in Figure 5b. Overall, these results thus demonstrate optical toggling between out-of-plane and in-plane correlated spin orientations in $CrPS_4$.

It is noteworthy that the power density of the control LED at the sample (140 $W/cm^2$) here is smaller than those reported in other recent studies of $CrPS_4$ (1-400 $kW/cm^2$).[15,17,37,44] Even accounting for $CrPS_4$ optical densities at different excitation wavelengths (405 nm here, 532 nm,[15,17,44] and 633 nm[17,37]), it is evident that common excitation power densities are sufficient to drive this SFT under an appropriate magnetic field, because only a small temperature jump is required.

In addition to demonstrating excellent *in situ* optical spin sensitivity of $Yb^{3+}$ dopants in $CrPS_4$, these findings represent the first demonstration of optical spin control in $CrPS_4$. Previously, magnetic switching in 2D magnets has been achieved by application of magnetic fields, electrical biases,[45,46] and external pressure.[47] Optically induced magnetic ordering[48] and demagnetization[40-43] have been demonstrated, and circularly polarized excitation has been used to invert spins in $CrI_3$,[49,50] but π/2 optical spin-reorientation (in-plane *vs* out-of-plane) is more challenging. The toggled π/2 spin-reorientation demonstrated here for $CrPS_4$ appears to be a unique example of light-driven correlated spin manipulation in 2D materials.

**Conclusion**

Doping $Yb^{3+}$ into $CrPS_4$ transforms the PL of this material by converting it to narrow sensitized *f-f* emission that is extremely sensitive to lattice spin. Site-selective PL measurements have identified $Yb^{3+}$ dopants substituting at the two distinct $Cr^{3+}$ lattice sites. The $Yb^{3+}$ PL shows pronounced exchange splittings that are highly dependent on magnetic correlations within the surrounding $CrPS_4$ lattice. Notably, the $CrPS_4$ metamagnetic spin-flop transition induces substantial changes in $Yb^{3+}$ PL energies and exchange splittings, allowing sensitive optical monitoring of lattice magnetism in this part of the magnetic phase diagram. This sensitivity has enabled the first demonstration of an optically driven spin-flop transition in a layered van der Waals magnet. The ability to control in-plane *vs* out-of-plane spin orientations using light suggests interesting opportunities for optical switching of tunneling magnetoresistance or other photo-spintronic functionality and it could enable dynamic patterning of spatially programmable spin textures with diffraction-limited feature sizes, offering novel paths toward reconfigurable quantum materials.

**Methods**

*Synthesis of $CrPS_4$ and $Yb^{3+}$-doped $CrPS_4$ crystals.* Crystals of the undoped $CrPS_4$ were grown by chemical vapor transport in a manner like that described previously,[51] and $Yb^{3+}$ doping was achieved by addition of ytterbium metal powder to the reaction mixture. A chromium chip (99.995%, lot MKCH4484) was purchased from Sigma Aldrich. The Cr chip was ground to a powder using a mortar and pestle. Red phosphorus powder (>97.0%, lot MKCS0319) was purchased from Sigma Aldrich. Sulfur powder (99.98%, lot STBL3899) was purchased from Sigma Aldrich. Ytterbium metal powder 40 mesh (99.9%) was purchased from BeanTown Chemical. All chemicals were used as received without further purification. In a typical synthesis, Yb metal (*x* mmol), Cr metal (2.6 – *x* mmol), P powder (2.6 mmol in P), and $S_8$ powder (1.3 mmol in S) were loaded (an *x*:1-*x*:1:4 mole ratio of the elements) into a quartz tube and sealed under an evacuated atmosphere. The quartz tubes were 10 cm long with inner and outer diameters of 14 and 16 mm, respectively. Sealed tubes were placed in an open-ended horizontal tube furnace with the starting materials in the hot zone set at 750 ℃ and the other end at a temperature of 650 ℃. Samples were heated for 5 days and then allowed to cool to room temperature over a period of ca. 6 hr. Once cooled, the tubes were cracked open to yield shiny dark plate-like crystals that had formed at the cold end. $Yb^{3+}$ doping was confirmed by inductively coupled plasma mass spectrometry (ICP-MS) using a PerkinElmer NexION 2000B. ICP-MS samples were digested in high-purity concentrated nitric acid, followed by dilution in ultrapure $H_2O$. $Yb^{3+}$ doping levels are reported as a percentage relative to the total cation contents, $[Yb^{3+}]/([Cr^{3+}]+[Yb^{3+}])$. Analytical $Yb^{3+}$ concentrations were ~50 times smaller than the nominal ytterbium concentrations used in the synthesis. The $Yb^{3+}$ concentrations in the flakes studied here were 0.20 ± 0.05% (Figures 1-3) and 0.020 ± 0.005% (Figures 4-5) with the uncertainty accounting for flake-to-flake variability.

*XRD measurements.* As-synthesized flakes were characterized by X-ray diffraction using a Bruker D8 Discover powder diffractometer with IμS microfocus X-ray source for Cu K$\alpha$ radiation (50 kV, 1 mA). Samples were placed onto crystalline silicon substrates and measured under ambient conditions.

*Raman scattering measurements.* Raman measurements were performed using a Renishaw inVia Raman microscope equipped with a 1200 grooves/mm grating and a Leica DMIRBE inverted optical microscope. A 532 nm laser was used as the excitation source. The collimated monochromatic beam was focused onto the sample through a Leica objective lens. Raman-scattered light was collected by the same objective, directed into the inVia spectrometer, dispersed by the diffraction grating, and focused on a CCD camera.

*Photoluminescence (PL) measurements.* Single crystals of the material were placed between two sapphire disks and loaded into a closed-cycle helium cryostat. Crystals were exfoliated prior to loading to ensure fresh surfaces, and measurement samples were typically 100-500 μm thick. Measurements were performed under high vacuum ($10^{-6}$ Pa). Sample emission was focused into a monochromator with a spectral bandwidth of 1 nm for the data in Figure 2a and 0.01 nm for all other PL data. The excitation source and collection axes are both normal to the crystal face. Spectra were collected using a $LN_2$-cooled silicon CCD camera. All PL spectra were corrected for instrument response. Photoexcitation was performed using either a 632.8 nm helium-neon laser, a tunable Ti:Sapphire laser, a 532 nm Nd:YAG laser, or a 405 nm LED, as specified in the text. Where specified, polarization scrambling was achieved by passing the excitation source and sample emission through fused silica depolarizers. Unless specifically indicated otherwise, crystal orientation was not controlled relative to the excitation polarization.

*Photoluminescence excitation (PLE) measurements.* PLE measurements were likewise conducted in a closed-cycle helium cryostat under high vacuum. Sample emission was focused into a monochromator with a spectral bandwidth of 5 nm and emission counts were monitored using a Hamamatsu InGaAs/InP NIR photomultiplier tube. Spectra were corrected for intensity of the excitation source, measured simultaneously using a calibrated Si diode.

*Magneto-PL measurements.* Individual sample flakes were placed between two quartz disks and loaded into a Cryo-Industries SMC-1650 OVT superconducting magneto-optical cryostat oriented in the Faraday configuration. The magnetic field was applied parallel to the crystallographic *c*-axis. Sample emission was collected along the magnetic field axis and focused into a fiber-optic cable. Sample emission was then passed through a monochromator with a spectral bandwidth of 0.25 nm and detected using a $LN_2$-cooled silicon CCD camera. Field-sweep measurements were performed by measuring sample emission every 1 s while continuously sweeping the magnetic field at a rate of 4 mT/s.

**Acknowledgments**

This research was primarily supported by the University of Washington Molecular Engineering Materials Center, a U.S. National Science Foundation Materials Research Science and Engineering Center (DMR-2308979). Part of this work was conducted at the Molecular Analysis Facility, which is supported in part by funds from the Molecular Engineering & Sciences Institute, the Clean Energy Institute, the National Science Foundation (NNCI-2025489 and NNCI-1542101).

**Contributions**

JTB and DRG conceived of this work. JTB led all aspects of experimentation with the support of other authors: KP (magneto-PL, data analysis), NJA (laser PLE, magneto-PL), FH (Raman analysis), TJS (crystal synthesis, ICP-MS, magneto-PL), RB (PL, data analysis). The manuscript was prepared by JTB and DRG in consultation with all other authors. DRG supervised the project. All authors read and commented upon the manuscript.

**Data Availability**

The original experimental data presented in this work will be made freely available at https://doi.org/10.5281/zenodo upon manuscript acceptance.

**Corresponding Authors**

Correspondence to Daniel R. Gamelin.

**Competing Interests**

The authors declare no competing interests.

**Supplementary Information**

Supplementary Figures S1-S17.

# Optical Spin Sensing and Metamagnetic Phase Control in the 2D Van der Waals Magnet $Yb^{3+}$-Doped $CrPS_4$

Jacob T. Baillie,[1] Kimo Pressler,[1] Nick J. Adams,[1]
Faris Horani,[1] Thom J. Snoeren,[1] Rémi Beaulac,[2] and Daniel R. Gamelin[1,*]

[1]*Department of Chemistry, University of Washington, Seattle, WA 98195*
[2]*Department of Chemistry, Swarthmore College, Swarthmore, PA 19081*
*Correspondence to: gamelin@uw.edu

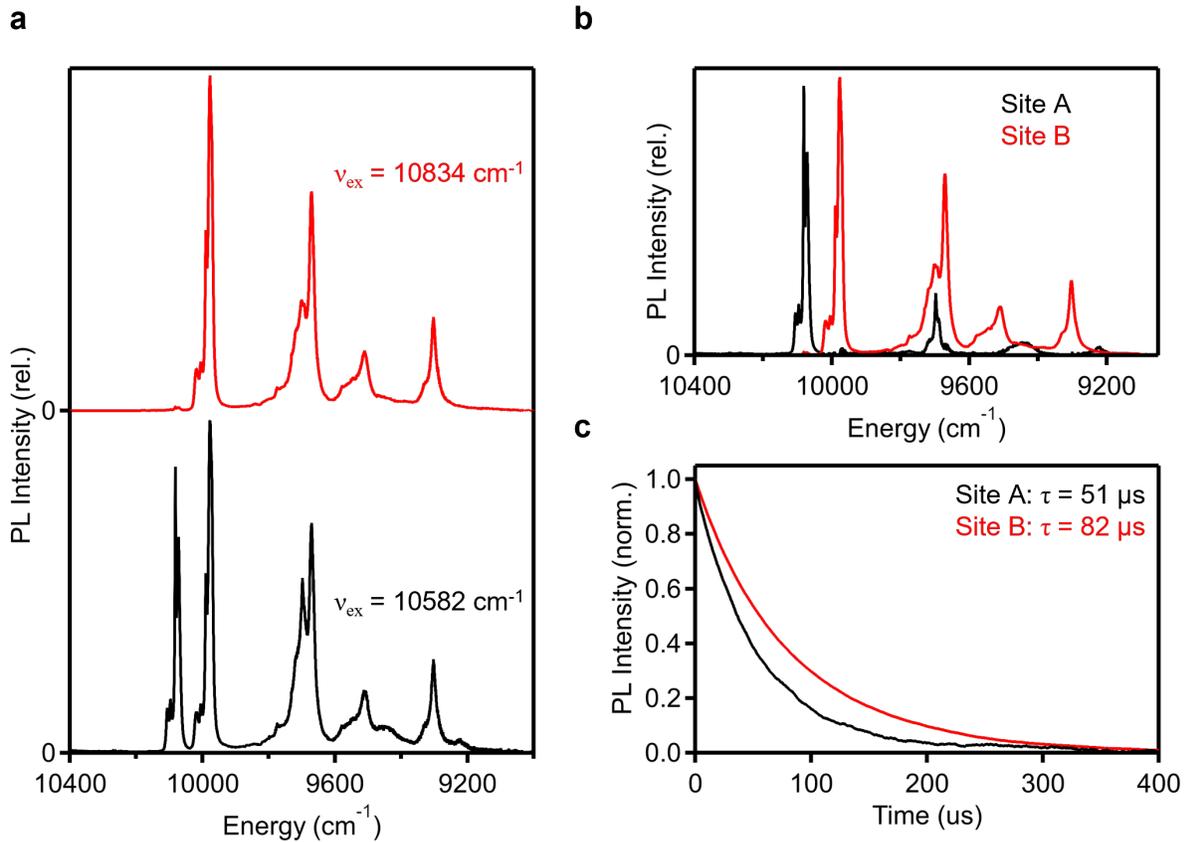

**Figure S1.** (a) Site-selective PL spectra of a 0.2% $Yb^{3+}$:$CrPS_4$ single crystal (same as Figure 3) measured at 20 K using a tunable, CW Ti:sapphire laser (~20 mW/cm$^2$). The red spectrum was obtained with excitation at 10834 cm$^{-1}$ (923 nm) and the black spectrum was obtained with excitation at 10582 cm$^{-1}$ (945 nm). (b) Individual PL spectra of $Yb^{3+}$ in Sites A and B, obtained by deconvolution of the spectra in panel (a). (c) Site-selective PL decay curves measured for the $Yb^{3+}$ 0' → 0 transitions of Sites A (black, 10080 cm$^{-1}$) and B (red, 9980 cm$^{-1}$) with 0.1 nm spectral bandwidth, showing that the two sites have different decay times. Lifetimes are determined by fitting the PL decay curves to a monoexponential decay function. Pulsed photoexcitation was performed using the 532 nm output of a frequency doubled Nd:YAG laser (50 Hz, 30 ps pulse duration, 4 nJ pulse energy) with a 100 cm$^{-1}$ bandwidth.

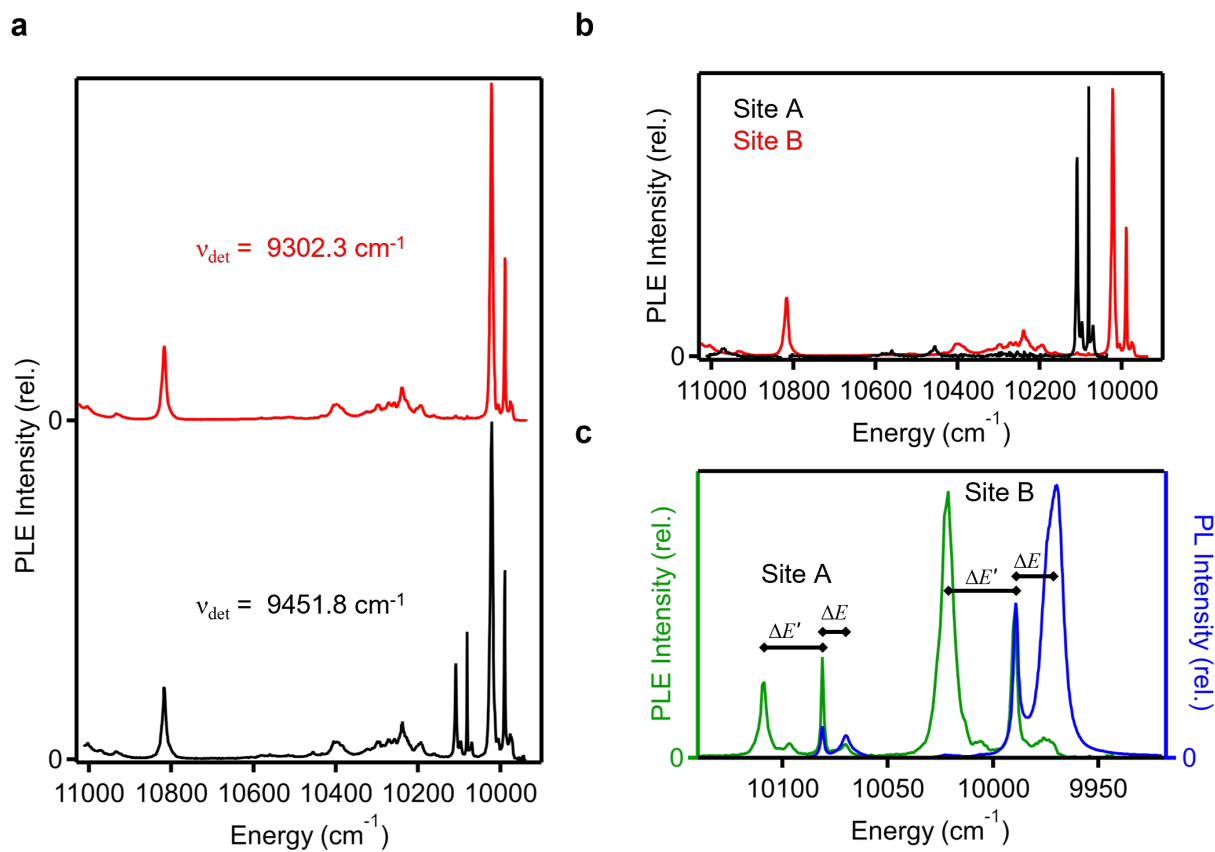

**Figure S2. (a)** Site-selective PLE spectra of a 0.2% $Yb^{3+}$:$CrPS_4$ single crystal (same as Figure 3) measured at 4 K. The red spectrum was obtained with detection at 9302.3 cm$^{-1}$ and the black spectrum was obtained with detection at 9451.8 cm$^{-1}$. **(b)** PLE spectra of the individual $Yb^{3+}$ Sites A and B, obtained by deconvolution from the spectra presented in panel (a). **(c)** PL and PLE spectra of the $Yb^{3+}$ 0' ↔ 0 transitions. PL was measured with unpolarized CW 15803 cm$^{-1}$ (632.8 nm) photoexcitation at 5 mW/cm$^2$. PLE was measured using a tunable, CW Ti:Sapphire laser (5-50 mW/cm$^2$) while monitoring PL at 9451.8 cm$^{-1}$, where both sites emit.

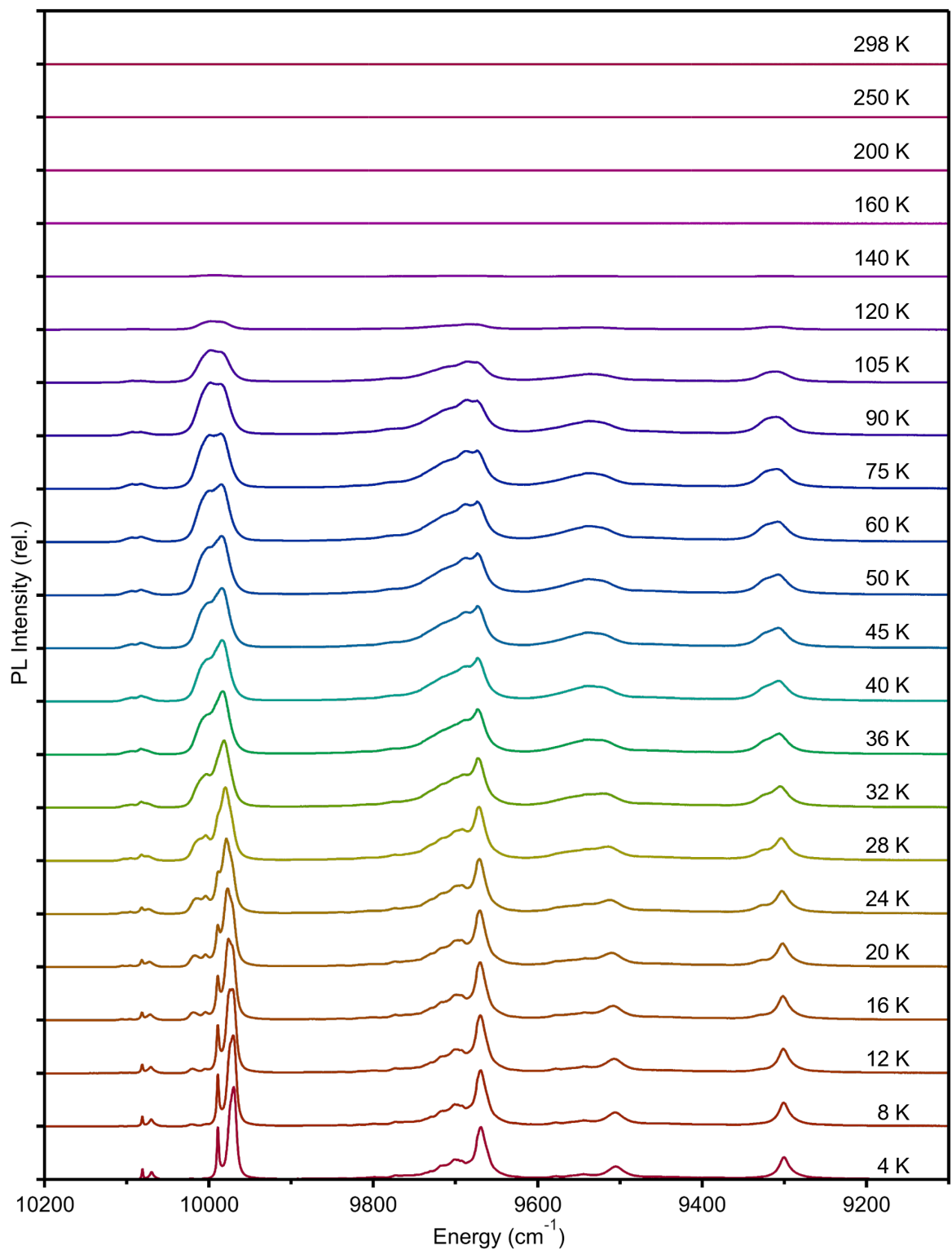

**Figure S3.** Variable-temperature PL spectra of a 0.2% $Yb^{3+}$-doped $CrPS_4$ single crystal (same as Figure 3) collected with unpolarized CW 632.8 nm photoexcitation at 5 $mW/cm^2$.

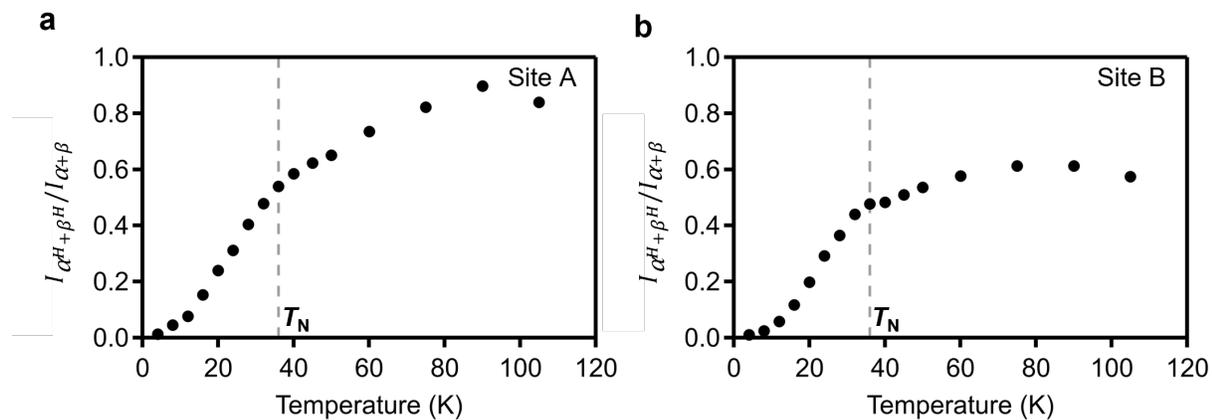

**Figure S4.** Temperature dependence of hot-band PL intensities from a 0.2% $Yb^{3+}$-doped $CrPS_4$ single crystal (same as Figure 3), plotted as $I_{\alpha^H+\beta^H}/I_{\alpha+\beta}$ vs temperature. $I_{\alpha+\beta}$ represents the integrated area of the combined peaks $\alpha$ and $\beta$ (cold bands) for a given $Yb^{3+}$ site (Site A or B), and $I_{\alpha^H+\beta^H}$ represents the integrated area of the combined peaks $\alpha^H$ and $\beta^H$ (hot bands) for the same site, as determined by the deconvolution described in Figure S6. **(a)** Site A data. **(b)** Site B data.

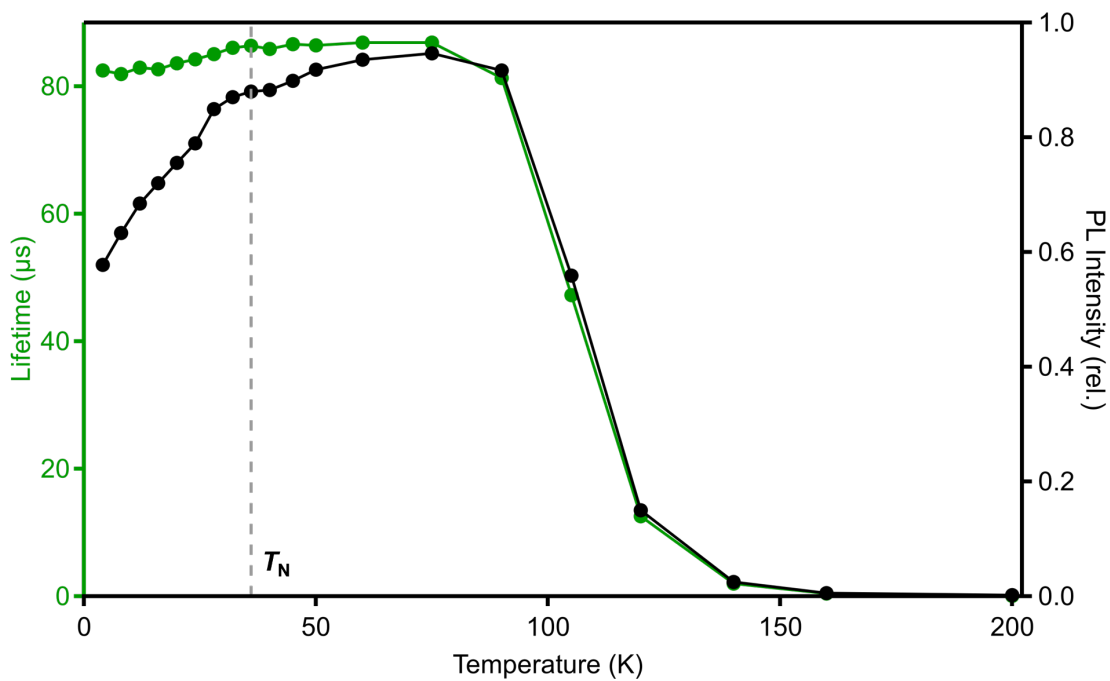

**Figure S5.** Temperature dependence of (black) the integrated intensity of the entire $Yb^{3+}$ PL spectrum ($10200 – 9200$ $cm^{-1}$) and (green) the PL decay time measured at $9970 \pm 50$ $cm^{-1}$ (Site B, both $\alpha$ and $\beta$ transitions) of a 0.2% $Yb^{3+}$-doped $CrPS_4$ single crystal (same as Figure 3). PL was measured with CW 632.8 nm photoexcitation at 5 mW/cm² exciting the $CrPS_4$ lattice. PL decay was collected using the 532 nm output of a frequency doubled Nd:YAG laser (50 Hz, 30 ps pulse duration, 4 nJ pulse energy) for excitation. All other major spectroscopic features assigned to Site B were also measured at 4 K and have very similar decay times with slight variation, likely due to partial overlap with Site A features. The divergence between $Yb^{3+}$ PL intensities and lifetimes as temperature is lowered below ~100K suggests a decreasing quantum efficiency of sensitization at lower temperatures.

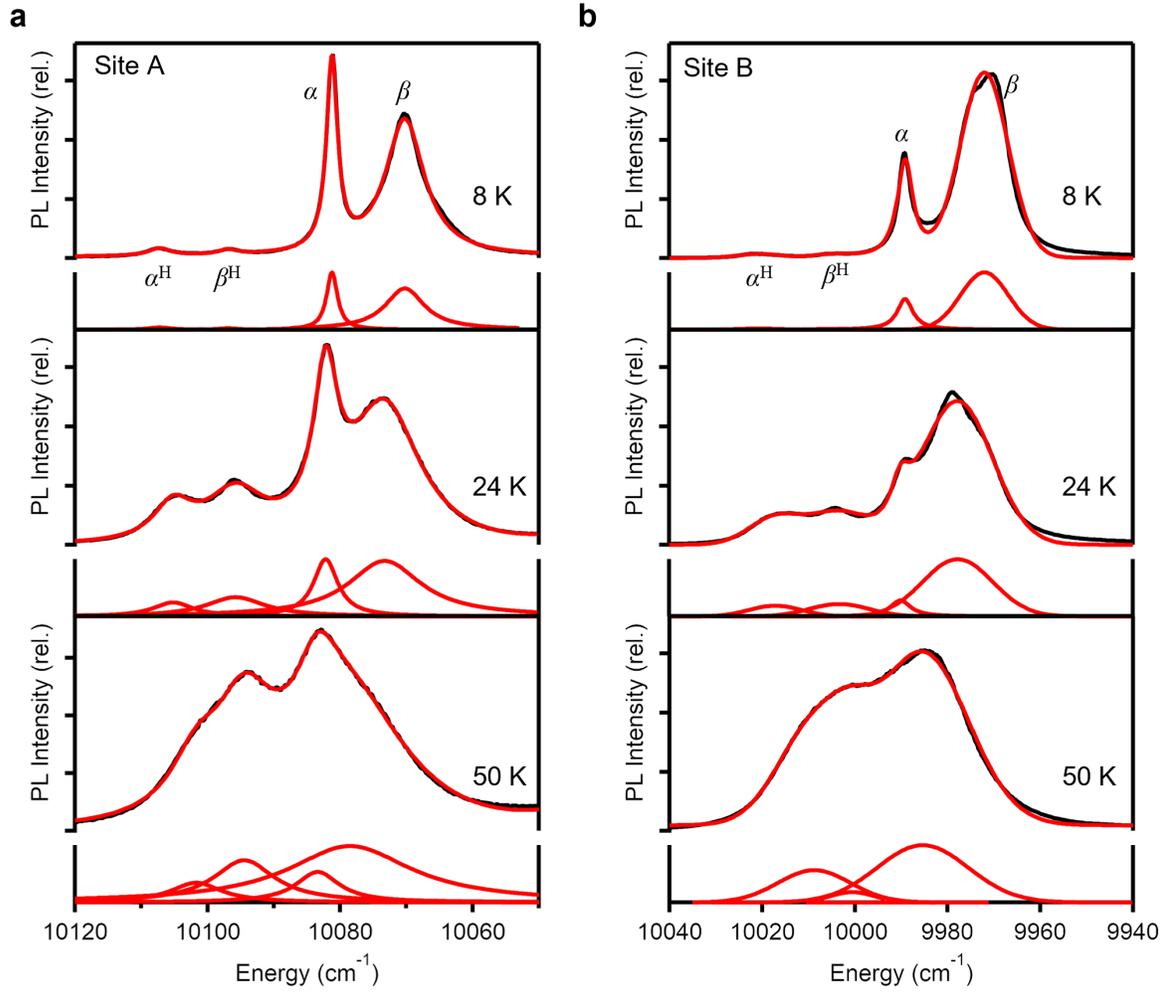

**Figure S6.** Spectral deconvolution of PL spectra of a 0.2% $Yb^{3+}$-doped $CrPS_4$ single crystal (same as Figure 3) in the region of the 0' → 0 transition for each $Yb^{3+}$ site, for three sample temperatures (8 K, 24 K, and 50 K). **(a)** Site A; deconvolution was performed by least-squares fitting to four Lorentzian peaks and a cubic baseline. **(b)** Site B; deconvolution was performed by least-squares fitting to three Gaussian peaks ($\beta$, $\alpha^H$, $\beta^H$) and one Lorentzian peak ($\alpha$), with a constant baseline. Peak positions from this fitting are used in the heat maps of Figure 3a,c and to determine $\Delta E_{exch}$ and $\Delta E'_{exch}$ energies in Figure 3b,d.

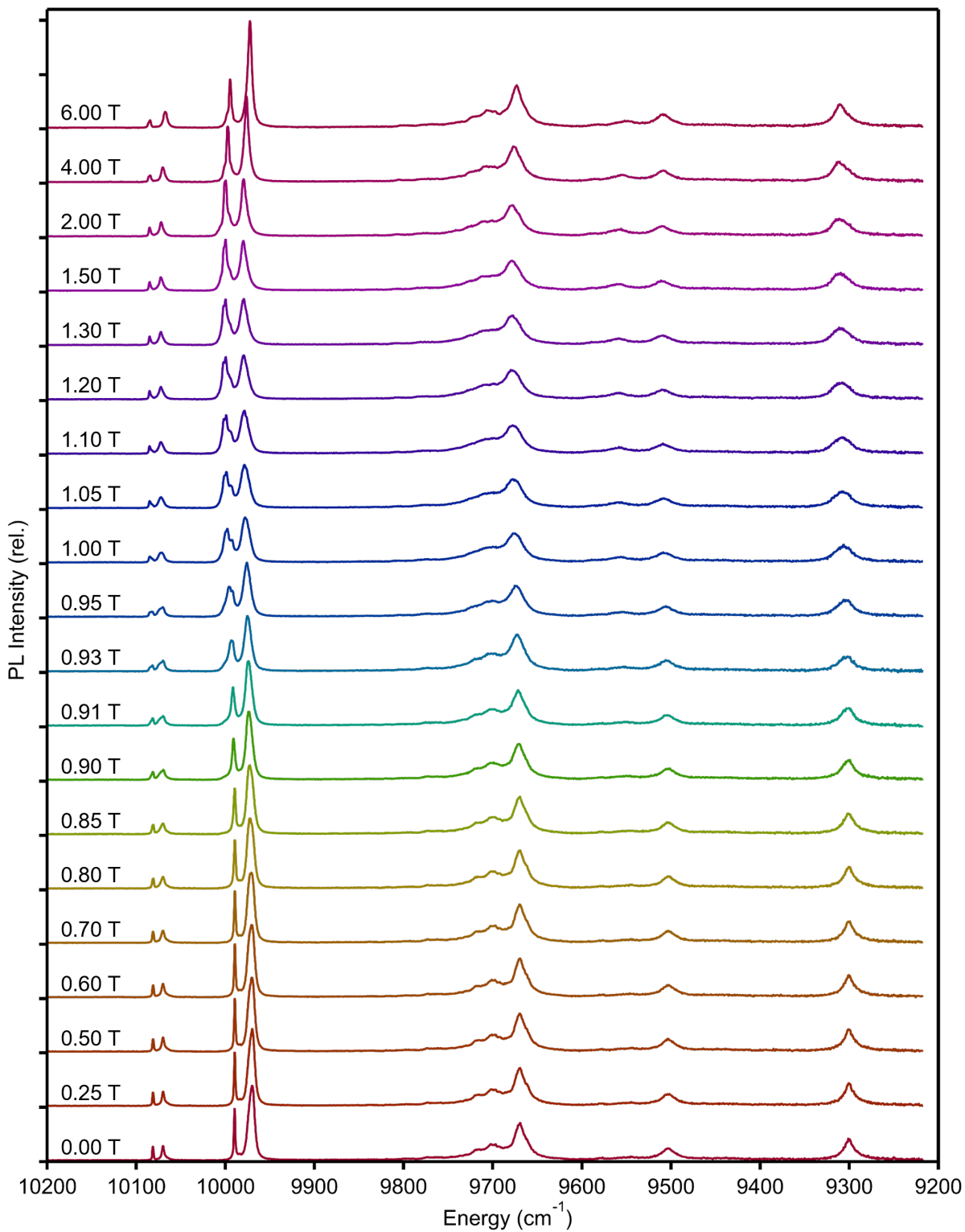

**Figure S7.** 4.6 K variable-field PL spectra of a 0.02% $Yb^{3+}$-doped $CrPS_4$ single crystal (same as Figure 4) measured with unpolarized CW 632.8 nm photoexcitation at 5 mW/cm$^2$. The magnetic field, incident excitation, and emission collection axes are all parallel to the crystallographic *c*-axis.

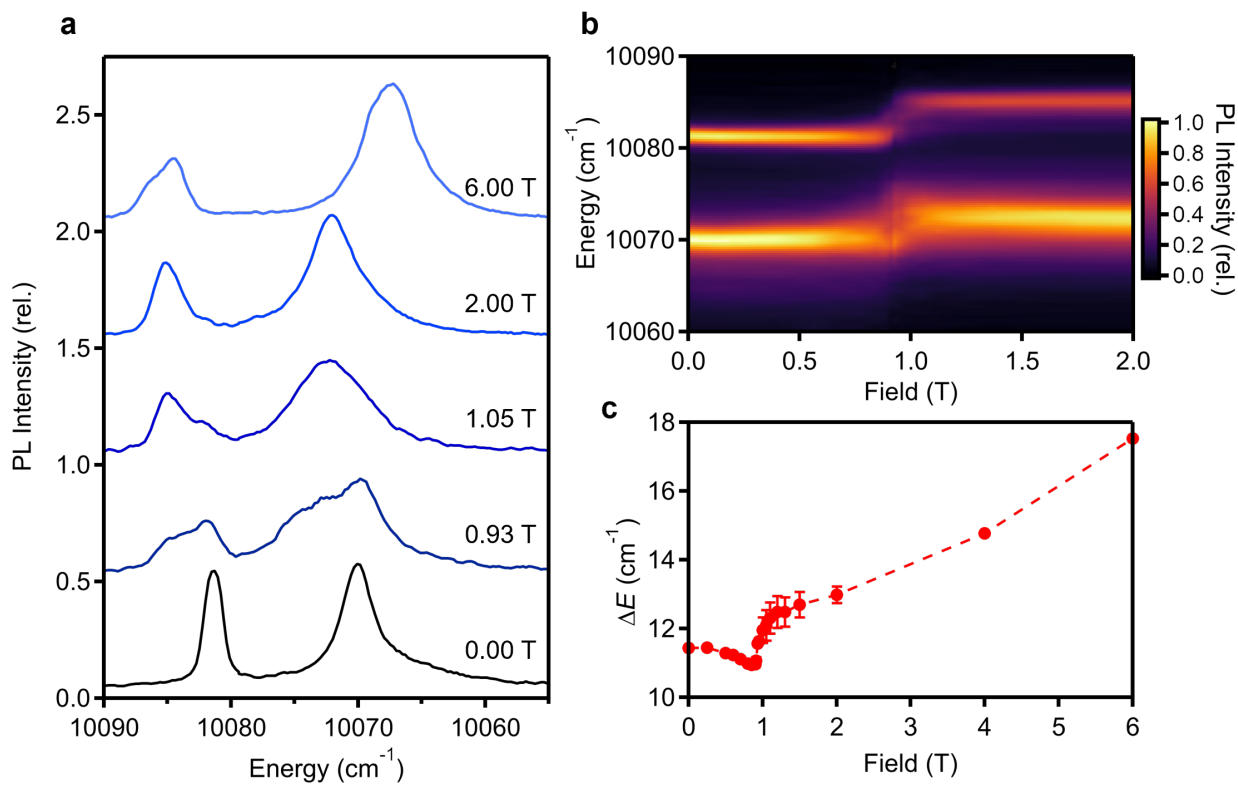

**Figure S8. (a)** Variable-field PL spectra in the region of the α and β peaks for Site A. **(b)** False-color plot of the PL intensities *vs* magnetic field strength in the region of the α and β peaks for Site A. **(c)** Plot of ΔE *vs* magnetic field strength for Site A. Determination of the peak energies is described in Figures S9 and S10. All panels are for an oriented 0.02% $Yb^{3+}$-doped $CrPS_4$ single crystal (same as Figure 4) measured at 4.6 K using CW 632.8 nm unpolarized photoexcitation at 5 mW/cm$^2$ with the incident optical and magnetic-field axes oriented parallel to the crystallographic *c*-axis.

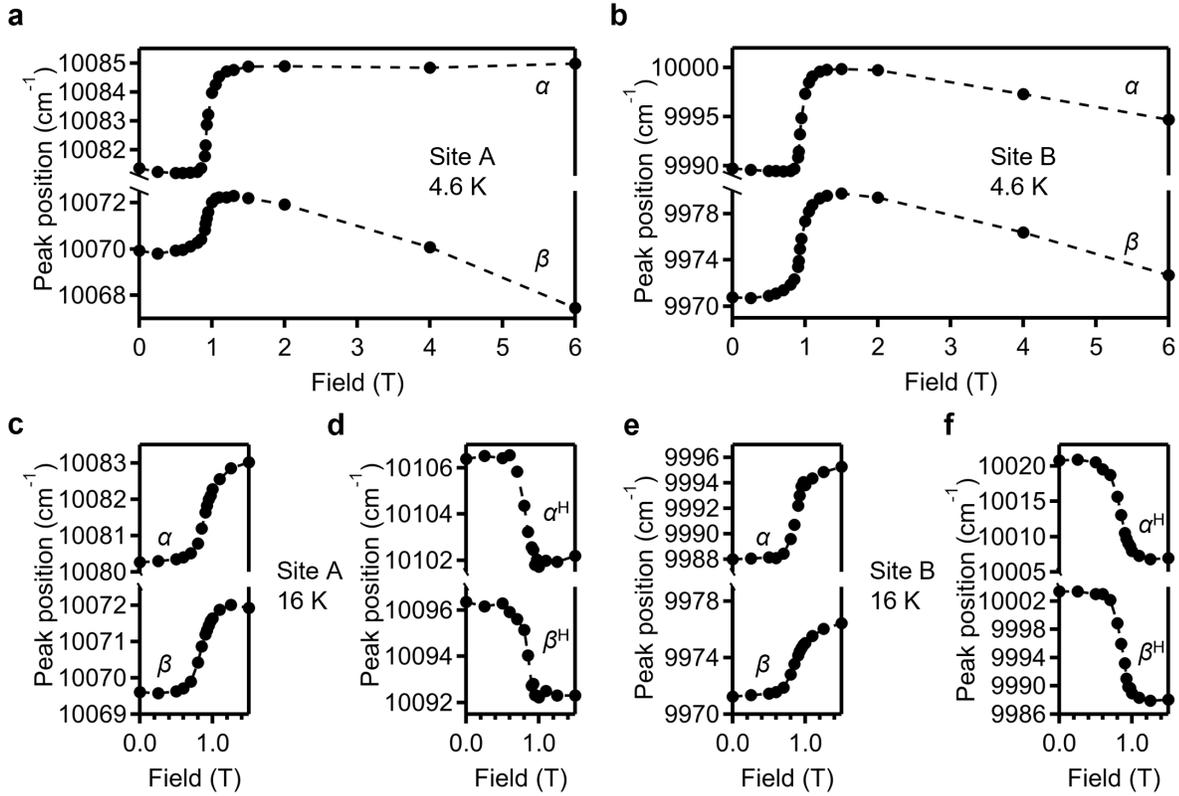

**Figure S9. (a)** Plots of *α* and *β* peak energies for Site A *vs **B*** at 4.6 K. **(b)** Plots of *α* and *β* peak energies for Site B *vs **B*** at 4.6 K. **(c)** Plots of *α* and *β* peak energies for Site A *vs **B*** at 16 K. **(d)** Plots of $\alpha^H$ and $\beta^H$ peak energies for Site A *vs **B*** at 16 K. **(c)** Plots of *α* and *β* peak energies for Site B *vs **B*** at 16 K. **(c)** Plots of $\alpha^H$ and $\beta^H$ peak energies for Site B *vs **B*** at 16 K. All data sets were collected from a 0.02% $Yb^{3+}$-doped $CrPS_4$ single crystal (same as Figure 4) using CW 632.8 nm unpolarized photoexcitation at 5 mW/cm$^2$. Peak energies were determined by spectral deconvolution (see Figure S10).

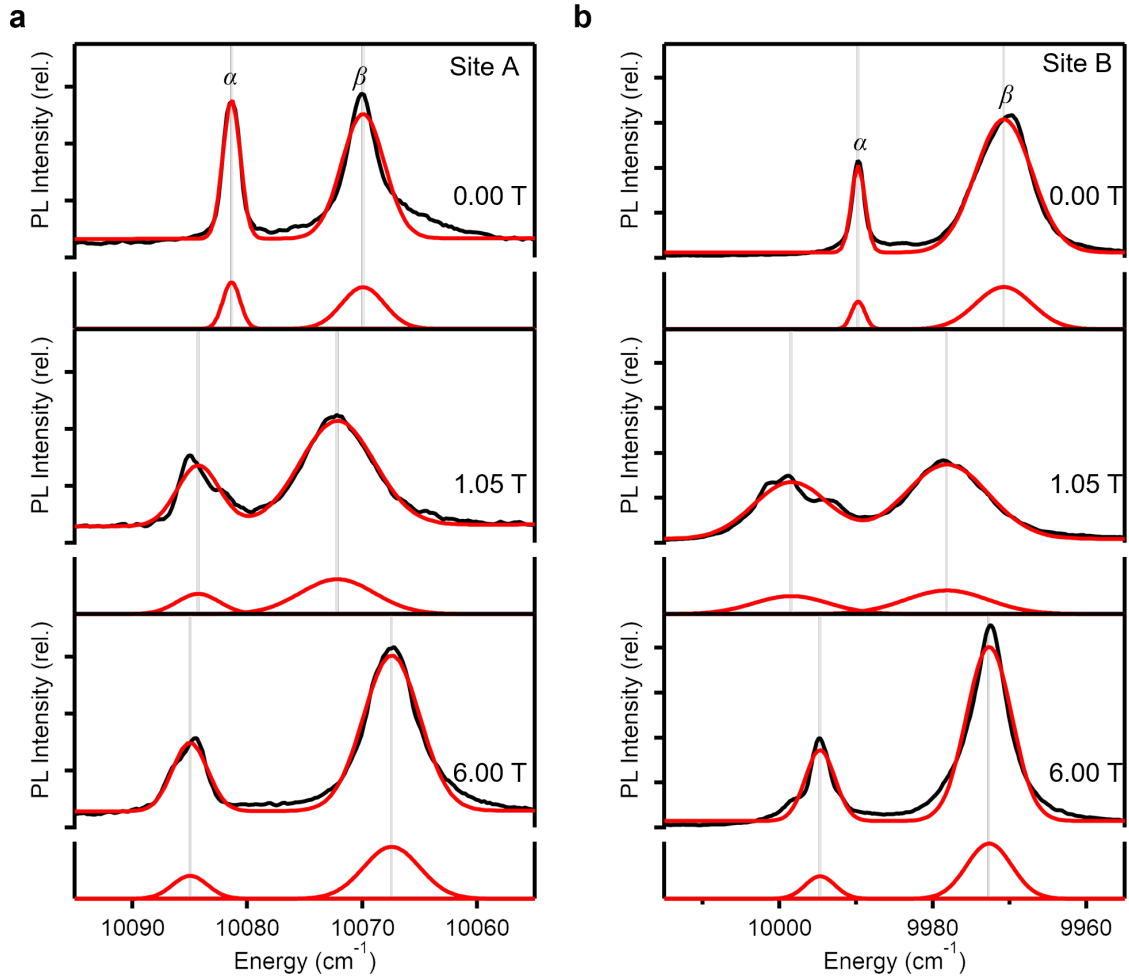

**Figure S10.** Spectral deconvolution of variable-field PL spectra of a 0.02% $Yb^{3+}$-doped $CrPS_4$ single crystal (same as Figure 4) in the region of the $\alpha$ and $\beta$ peaks for Sites A and B at 4.6 K. **(a)** Site A: deconvolution is performed by least-squares fitting to a linear baseline and two Gaussian peaks. **(b)** Site B: deconvolution is performed by least-squares fitting to a linear baseline and two Gaussian peaks. Peak positions obtained from this fitting are plotted in Figure S9 and used to determine the $\Delta E$ values plotted in Figures 4c and S8. The same analysis (data not shown) was performed for the data collected at 16 K but with four Gaussian peaks and was used to plot the peak position and $\Delta E$ and $\Delta E'$ splittings in Figures S9 and S11.

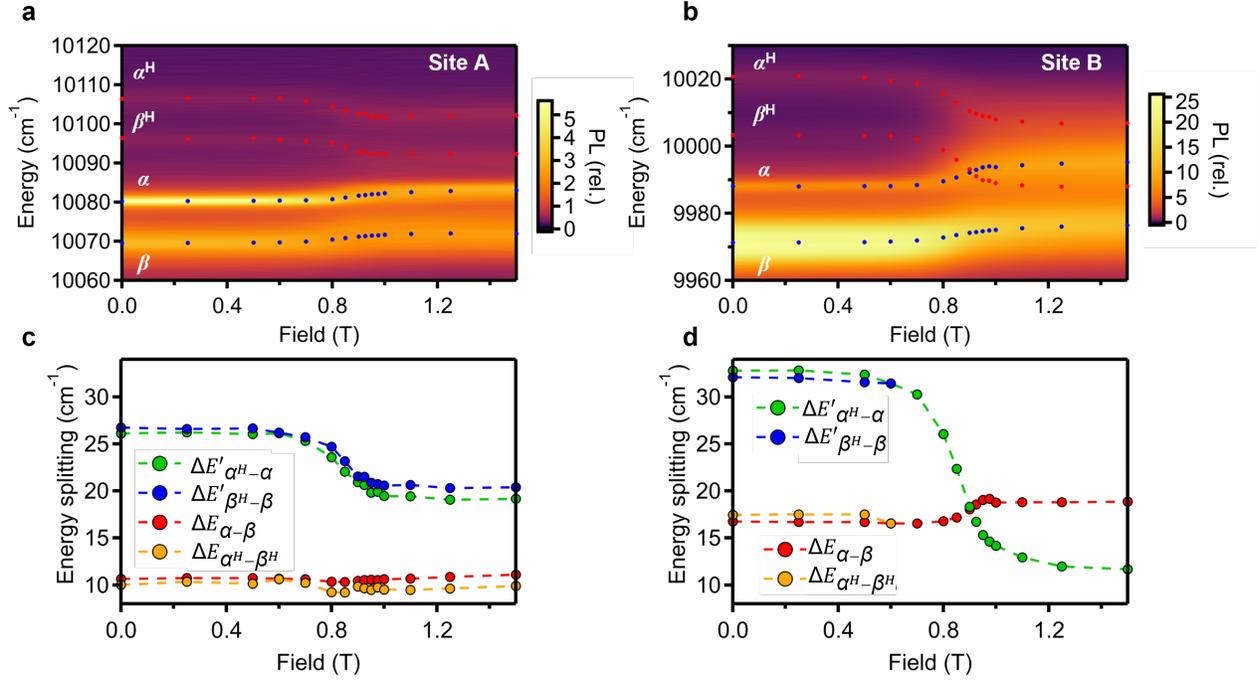

**Figure S11. (a)** False-color plot of the 16 K PL intensities *vs* magnetic field strength in the region of the 0' → 0 peaks of $Yb^{3+}$ in Site A. The circular data points plot the energies of the $\alpha^H$ and $\beta^H$ peaks in red and of the $\alpha$ and $\beta$ peaks in blue, as determined by the spectral deconvolution described in Figure S10. **(b)** Same as panel (a) but at Site B. **(c)** Magnetic field dependence of $\Delta E$ and $\Delta E'$ for Site A, determined from the data in panel (a) as described in the main text (Figure 3) and in Figure S12. **(d)** Same as panel (c) but for Site B. All data were collected from a 0.02% $Yb^{3+}$-doped $CrPS_4$ single crystal (same as Figure 4), using unpolarized CW 632.8 nm photoexcitation at 5 mW/cm$^2$.

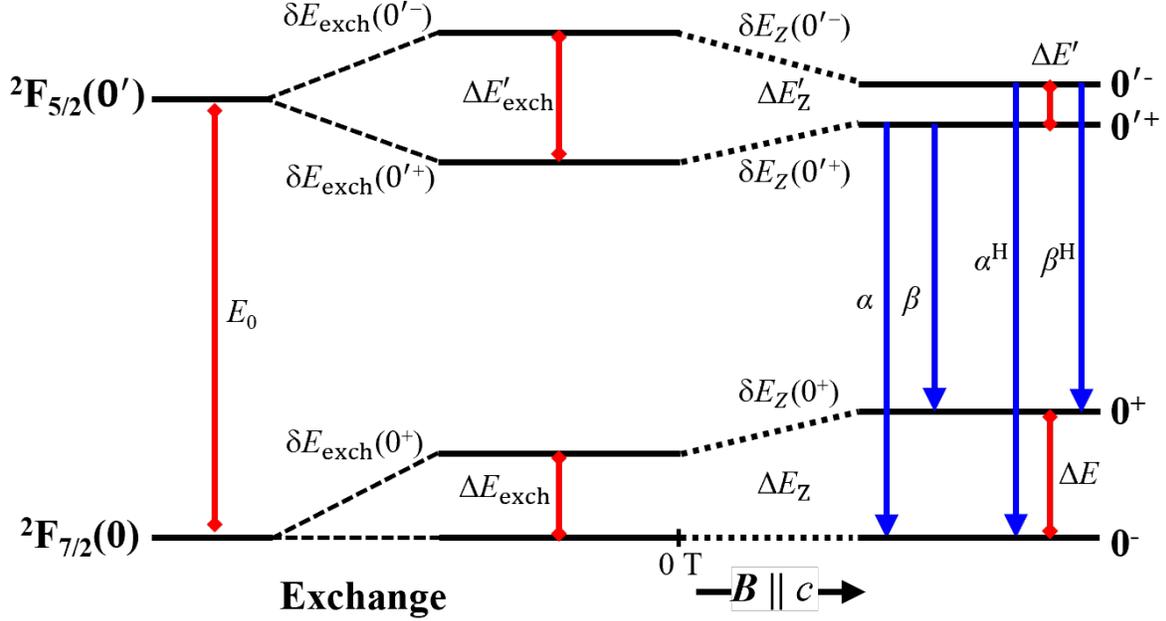

$$E(0^-) \equiv 0 \text{ cm}^{-1} \tag{S1}$$
$$E(0^+) = E(\alpha) - E(\beta) = \delta E_{exch} + \delta E_Z(0^+) \tag{S2}$$
$$E(0'^+) = E(\alpha) = E_0 + \delta E_{exch}(0'^+) + \delta E_Z(0'^+) \tag{S3}$$
$$E(0'^-) = E(\alpha^H) = E_0 + \delta E_{exch}(0'^-) + \delta E_Z(0'^-) \tag{S4}$$

$$\Delta E_{exch} = \delta E_{exch}(0^+) - \delta E_{exch}(0^-) \equiv \delta E_{exch}(0^+) \tag{S5}$$
$$\Delta E_Z = \delta E_Z(0^+) - \delta E_Z(0^-) \equiv \delta E_Z(0^+) \tag{S6}$$
$$\Delta E = \Delta E_{exch} + \Delta E_Z \tag{S7}$$

$$\Delta E'_{exch} = \delta E_{exch}(0'^-) - \delta E_{exch}(0'^+) \tag{S8}$$
$$\Delta E'_Z = \delta E_Z(0'^-) - \delta E_Z(0'^+) \tag{S9}$$
$$\Delta E' = \Delta E'_{exch} + \Delta E'_Z \tag{S10}$$

**Figure S12.** Energy-level diagram showing PL transitions between exchange-split $Yb^{3+}$ states. $E(0^-)$ is defined as 0 cm$^{-1}$ (eq S1), and $E(0^+)$, $E(0'^+)$, and $E(0'^-)$ are defined by eqs S2-4 as illustrated in the diagram. $E_0$ represents the combination of spin-orbit and crystal-field energies that place the $^2F_{5/2}(0')$ excited state ca. 10,000 cm$^{-1}$ above the $^2F_{7/2}(0)$ ground state in the absence of magnetic-exchange or Zeeman contributions to these energies. $\Delta E_{exch}$ represents the magnetic-exchange splitting of the $^2F_{7/2}(0)$ ground state, and $\Delta E'_{exch}$ represents the magnetic-exchange splitting of the $^2F_{5/2}(0')$ excited state. $\delta E_{exch}$ denotes the change in the energy of a given state relative to the ground state due to magnetic-exchange coupling. $\delta E_Z$ denotes the change in the energy of a given state relative to the ground state due to an external magnetic field oriented perpendicular to the sample plane. The total splitting of the $^2F_{7/2}(0)$ state resulting from the combined effects of the exchange splitting ($\Delta E_{exch}$) and the Zeeman splitting ($\Delta E_Z$) is $\Delta E$ (eq S7). The analogous variables for $^2F_{7/2}(0')$ ($\Delta E'_{exch}$, $\Delta E'_Z$, $\Delta E'$) are defined similarly (eq S8-10).

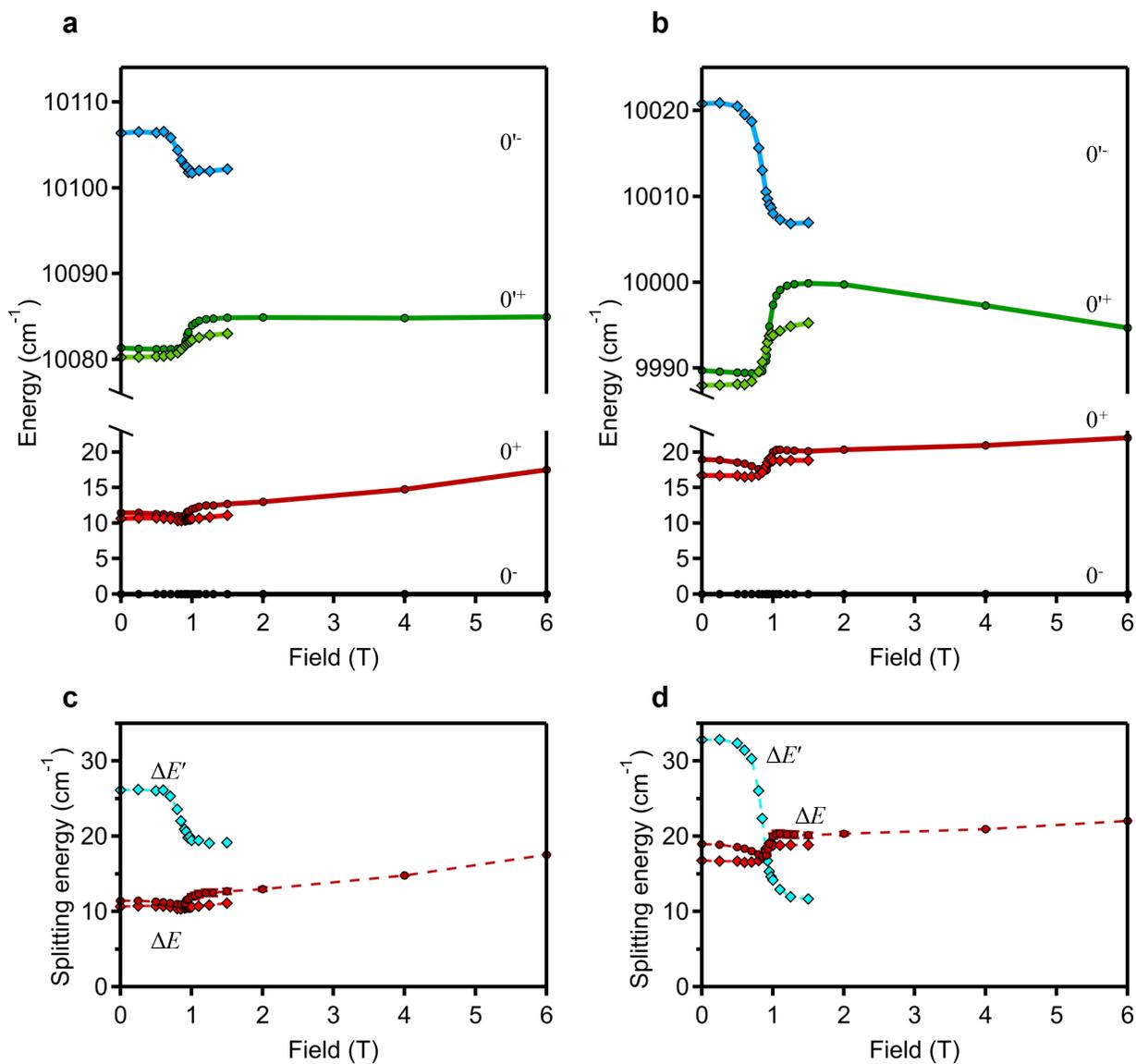

**Figure S13. (a)** Magnetic-field dependence of the energies of the $Yb^{3+}$ $0^-$, $0^+$, $0'^+$, and $0'^-$ states for Site A, measured from PL spectra collected at 4.6 K (circles, 0 – 6 T) and 16 K (diamonds, 0 – 1.5 T). **(b)** Same as panel (a) but for Site B. **(c)** $\Delta E$ and $\Delta E'$ values plotted *vs* magnetic field strength for $Yb^{3+}$ in Site A, determined from the data in panel (a) following Figure S12. **(d)** Same as panel (c) but for Site B. All data were collected from a 0.02% $Yb^{3+}$-doped $CrPS_4$ single crystal (same as Figure 4).

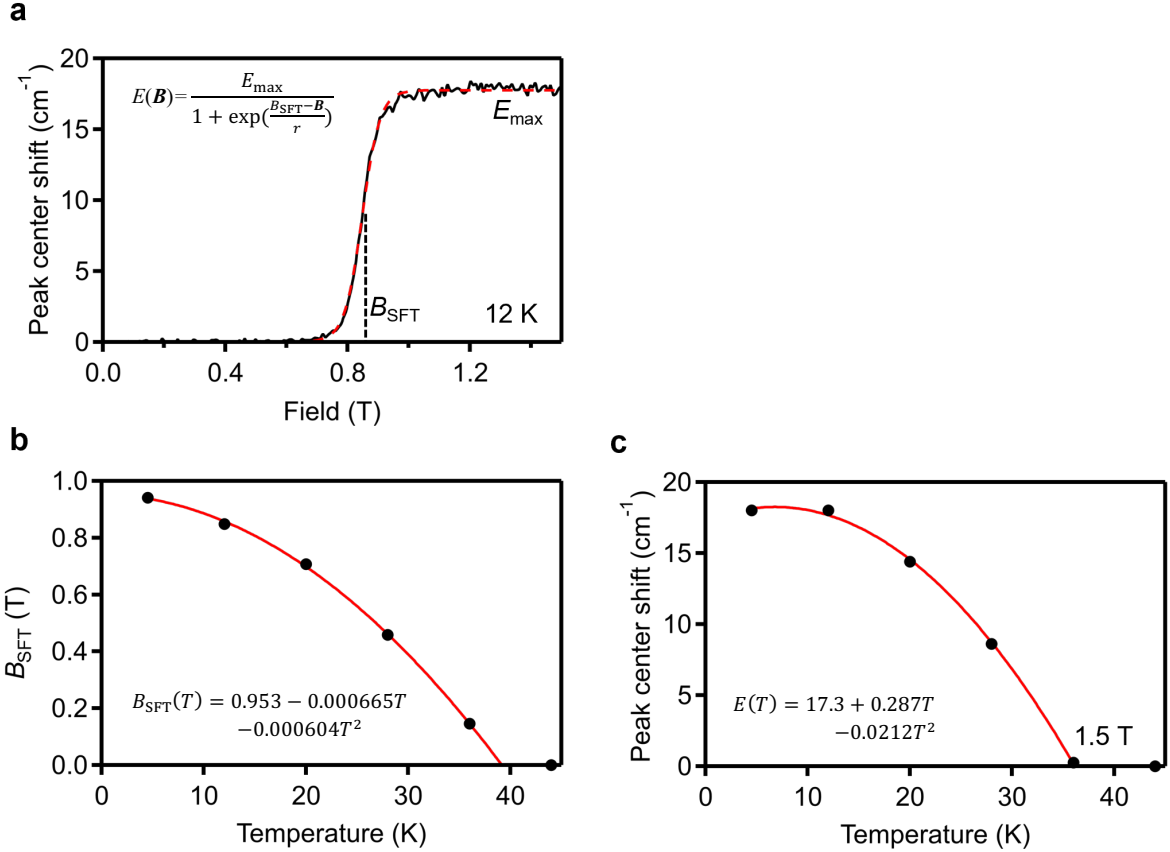

**Figure S14. (a)** Recast version of the 12 K data in Figure 5a showing the fitting used to determine $B_{SFT}$. The equation of the red dashed line is shown in the inset, where $E$ denotes the peak center shift, $E_{max}$ is the maximum shift, $r$ is a constant that defines the steepness of the sigmoid and is inversely related to the maximum slope, and $B_{SFT}$ defines the field at which 0.5 $E_{max}$ is reached. **(b)** Plot of the spin-flop field, $B_{SFT}$, from each plot in Figure 5a of the main text, determined by fitting the data to a sigmoidal function as described in panel (a). $E$ represents the shift in the center of spectral intensity for the combined 0' → 0 PL peaks and is used to find the field at which the midway point of the SFT occurs ($B_{SFT}$). Plotting the field at which the maximum change in PL energy occurs (obtained from the first derivatives of the data in Figure 5a of the main text) for each temperature gives the same result. The red curve shows a polynomial best fit to $B_{SFT}$ data collected between 4.6 and 36 K, yielding the numerical results shown in the inset (units of coefficients are implicit). **(c)** Plot of the PL energy measured at 1.5 T relative to its value at 0 T and the same temperature for each experimental temperature shown in Figure 5a. The red curve shows a polynomial fit to the data between 4.6 and 36 K, yielding the numerical results shown in the inset (units of coefficients are implicit). The fitted curves in panels b and c were used to generate the interpolated points in the heatmap of Figure 5b.

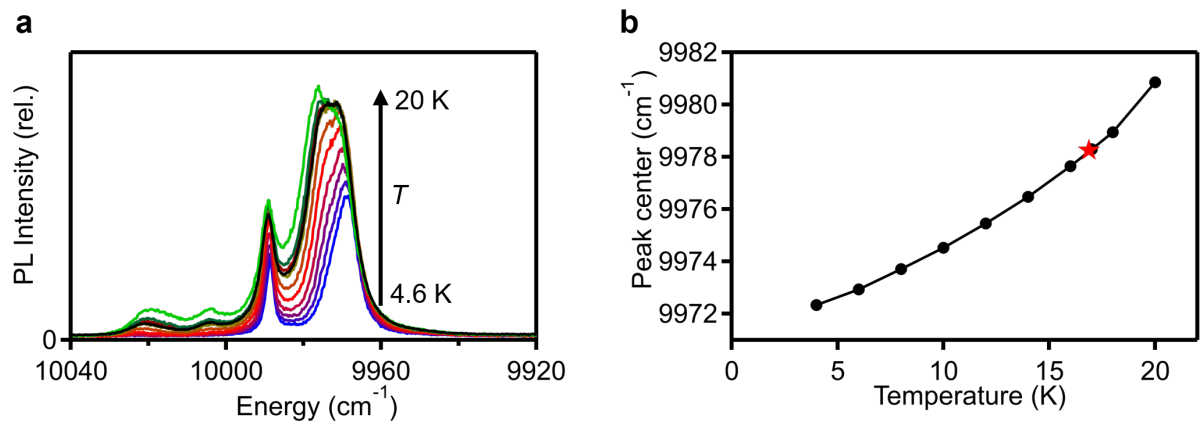

**Figure S15. (a)** Variable-temperature $Yb^{3+}$ PL spectra for $Yb^{3+}$ in Site B, measured in the region of the 0' → 0 transition from 4.6 to 20 K using unpolarized CW 632.8 nm photoexcitation at 5 mW/cm$^2$. The black spectrum was measured at 4.6 K with photothermal heating from an unpolarized CW 405 nm "control" LED operating at 140 W/cm$^2$. **(b)** Temperature dependence of the average PL energies in panel (a), obtained by integrating the PL intensities between 9940 – 10040 cm$^{-1}$. The star indicates the average PL energy for the photothermal (black) spectrum in panel (a), from which an internal temperature of 17 K is deduced. All data were collected from a 0.02% $Yb^{3+}$-doped $CrPS_4$ single crystal (same as Figure 4).

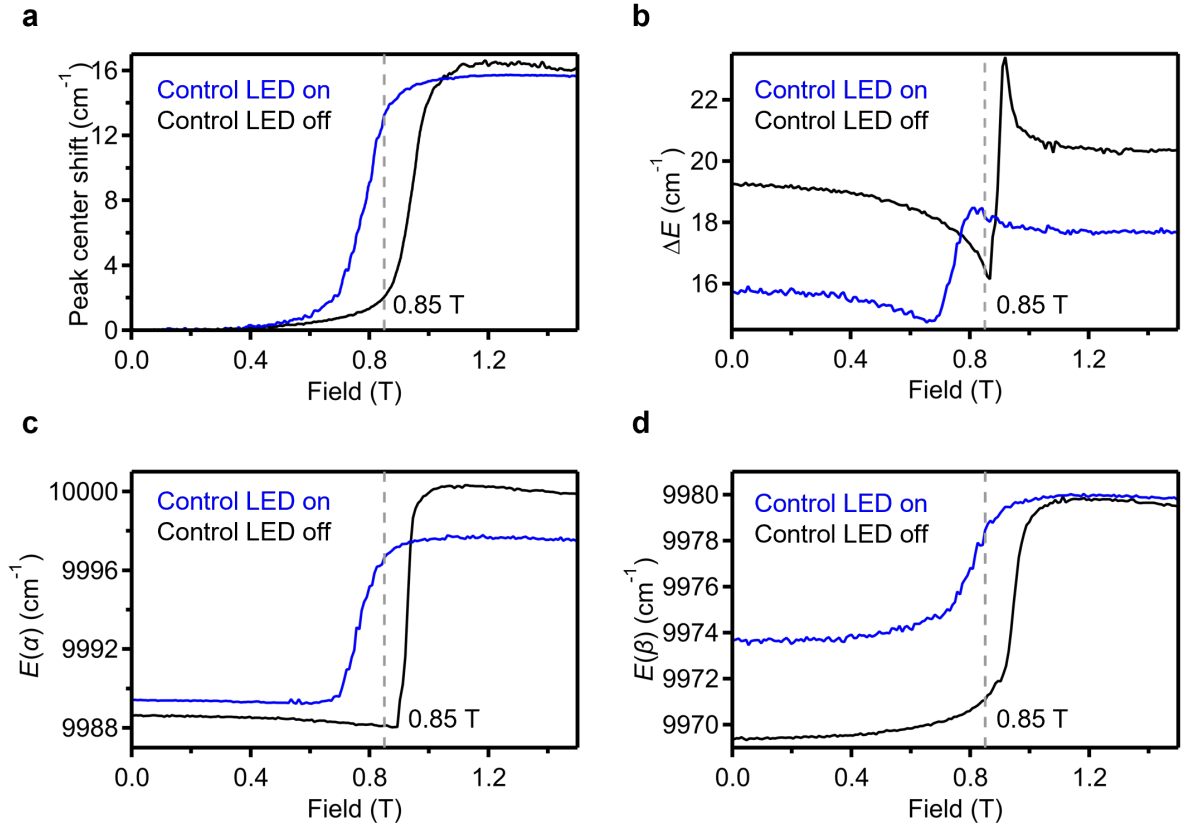

**Figure S16. (a)** Magnetic-field dependence of the centers of gravity for the combined $\alpha$ and $\beta$ PL peaks for $Yb^{3+}$ in Site B, measured under low (black, 632.8 nm, 5 mW/cm$^2$) and high (blue, 405 nm, 140 W/cm$^2$ and 632.8 nm, 5 mW/cm$^2$) unpolarized photoexcitation power densities at 4.6 K (*i.e.*, with or without photothermal heating). The dashed line indicates a field strength of 0.85 T. The peak-center shift was determined from average spectral intensities by integration over the range 9940 – 10040 cm$^{-1}$. **(b)** Same as panel (a), but for the PL splitting energy, $\Delta E = E(\alpha) - E(\beta)$ (*cf.* Fig. S12). **(c)** Same as panel (a), but for the energy of the $\alpha$ peak. **(d)** Same as panel (a), but for the energy of the $\beta$ peak. All data were collected from a 0.02% $Yb^{3+}$-doped $CrPS_4$ single crystal (same as Figure 4).

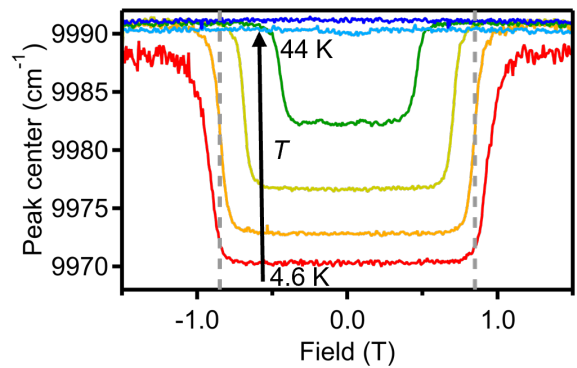

**Figure S17.** Recast version of Figure 5a from the main text, showing the centers of PL intensity for the combined *α* and *β* peaks *vs* magnetic-field strength, measured from 4.6 to 44 K for $Yb^{3+}$ in Site B. The dashed lines indicate magnetic-field strengths of -0.85 and 0.85 T. All data were collected from a 0.02% $Yb^{3+}$-doped $CrPS_4$ single crystal (same as Figure 4) using unpolarized CW 632.8 nm photoexcitation at 5 mW/cm$^2$.